\documentclass[preprint,tightenlines,aps,prd,showpacs,nofootinbib]{revtex4}
\usepackage{graphicx}
\usepackage{amsmath}
\usepackage{amssymb}
\usepackage{bm}

\begin{document}

\title{\mbox{}\\[10pt]
Weakly-bound Hadronic Molecule\\
near a 3-body Threshold}

\author{Eric Braaten and Meng Lu}
\affiliation{
Physics Department, Ohio State University,
Columbus, Ohio 43210, USA}
\author{Jungil Lee}
\affiliation{
Department of Physics, Korea University,
Seoul 136-701, Korea}

\date{\today}
\begin{abstract}
The $X(3872)$ seems to be a loosely-bound hadronic molecule whose
constituents are two charm mesons. A novel feature of this molecule
is that the mass difference of the constituents is close to the mass
of a lighter meson that can be exchanged between them, namely the pion.
We analyze this feature in a simple model with spin-0 mesons only.
Various observables are calculated to next-to-leading order in
the interaction strength of the exchanged meson.
Renormalization requires summing a geometric series
of next-to-leading order corrections.
The dependence of observables on the ultraviolet cutoff can be
removed by renormalizations of the mass of the heaviest meson,
the coupling constant for the contact interaction
between the heavy mesons, and short-distance
coefficients in the operator product expansion.
The next-to-leading order correction has
an unphysical infrared divergence at the threshold of the two heavier
mesons that can be eliminated by a further resummation that
takes into account the nonzero width of the heaviest meson.
\end{abstract}

\pacs{12.38.-t, 12.39.St, 13.20.Gd, 14.40.Gx}


\maketitle


\section{Introduction}

The $X(3872)$ is a new hadronic resonance discovered in 2003
by the Belle Collaboration \cite{Choi:2003ue}
and subsequently confirmed by the CDF, Babar, and D0 Collaborations
\cite{Acosta:2003zx,Abazov:2004kp,Aubert:2004ns}.
All the properties of the $X(3872)$ that have been measured
thus far are compatible with its identification as a
weakly-bound molecule whose constituents are a superposition
of the charm meson pairs $D^{*0} \bar D^0$ and $D^0 \bar D^{*0}$
\cite{Tornqvist:2003na,Close:2003sg,Pakvasa:2003ea,
Voloshin:2003nt,Wong:2003xk,Braaten:2003he,Swanson:2003tb,Braaten:2004rw,
Tornqvist:2004qy,Braaten:2004fk,Swanson:2004pp,Voloshin:2004mh,
Braaten:2004ai,Braaten:2005jj,AlFiky:2005jd,Braaten:2005ai,Suzuki:2005ha,
Swanson:2006st,Braaten:2006sy,AbdEl-Hady:2006ds,Colangelo:2007ph}.

Because it is so weakly bound, the $X(3872)$ has many features
in common with the deuteron, which is a
weakly-bound baryonic molecule consisting of a proton and a neutron.
Their binding energies are both small
compared to the natural energy scale
associated with pion exchange: $m_\pi^2 / (2M_{12})$,
where $M_{12}$ is the reduced mass of the two constituents.
The binding energy 2.2~MeV of the deuteron is small
compared to the natural scale of about 20~MeV.
After taking into account a recent precision measurement
of the $D^0$ mass by the CLEO Collaboration \cite{Cawlfield:2007dw},
the difference between the measured mass of the $X(3872)$ and the
$D^{*0} \bar D^0$ threshold is
\begin{equation}
M_X - (M_1 + M_2) = -0.6 \pm 0.6 \ {\rm MeV} .
\label{MX-CLEO}
\end{equation}
This is small compared to the natural energy scale of about 10~MeV.
The Belle Collaboration has set an upper bound on the width of the $X$
\cite{Choi:2003ue}:
\begin{equation}
\Gamma_X < 2.3 \ {\rm MeV} \hspace{1cm} (90\% \ {\rm C.L.}).
\label{GammaX-Belle}
\end{equation}
This is also small compared to the natural energy scale.
The small width is most easily understood
if the $X$ is below the $D^{*0} \bar D^0$ threshold.

Because the binding energies of the deuteron and the $X(3872)$
are small compared to the natural energy scales, they have universal
properties that are determined by the large scattering length $a$
of the constituents \cite{Braaten:2003he}.
The universality of few-body systems with a large scattering length
has many applications in atomic, nuclear, and particle physics
\cite{Braaten:2004rn}.
The universal features of the $X(3872)$ were first exploited by
Voloshin to describe its decays into $D^0 \bar D^0 \pi^0$ and
$D^0 \bar D^0 \gamma$, which can proceed through decay of the
constituent $D^{*0}$ or $\bar D^{*0}$ \cite{Voloshin:2003nt}.
Universality has also been applied to the
production process $B \to KX$ \cite{Braaten:2004fk,Braaten:2004ai},
to the line shape of the $X$ \cite{Braaten:2005jj}, and
to decays of $X$ into $J/\psi$ and pions \cite{Braaten:2005ai}.
These applications rely on factorization formulas that separate
the length scale $a$ from all the shorter distance scales of QCD
\cite{Braaten:2005jj}.  The factorization formulas can be derived
using the operator product expansion for a low-energy
effective field theory \cite{Braaten:2006sy}.

There is one feature of the $X(3872)$ that is very different
from the deuteron.  The mass difference between the
constituents $D^{*0}$ and $\bar D^0$, $M_2-M_1 = 142.12 \pm 0.07$ MeV,
is very close to the $\pi^0$ mass: $m=134.98$ MeV.
Thus the $D^{*0}$ mass is very close to the
$D^0 \pi^0$ threshold: $M_2-M_1 - m = 7.14 \pm 0.07$ MeV.
This is important because the pion is the lightest meson
that can be exchanged between the $D^{*0}$ and $\bar D^0$.
The small energy gap between the $D^{*0}$ mass and the $D^0 \pi^0$
threshold makes it easy for the $D^{*0}$
to make a transition to $D^0 \pi^0$.
An important consequence is that the $X$ could have
a substantial 3-body component $D^0 \bar D^0 \pi^0$.
Suzuki has argued that the near equality of $M_2-M_1$ and $m$
implies that the pion-exchange interaction is too weak to
bind the $X$ \cite{Suzuki:2005ha},
but this conclusion has been criticized \cite{Swanson:2006st}.

An analysis of the effect of the near equality $M_2-M_1 \approx m$
in the case of the $X(3872)$ is complicated by other effects
that may be comparable in importance.  There are other nearby
two-body and three-body thresholds.  The thresholds for the charged
mesons $D^{*+} D^-$ and $D^+ D^{*-}$ are higher than the $D^{*0} \bar D^0$
threshold by $8.08 \pm 0.12$ MeV.  The $D^0 D^- \pi^+$ and
$D^+ \bar D^0 \pi^-$ thresholds are higher than the
$D^{*0} \bar D^0$ threshold by only $2.23 \pm 0.12$ MeV.
There are several other features that complicate the analysis of
the effects of pions on the $X$.
The constituent $D^*$'s are spin-1 particles.
The charge conjugation $C = +$ requires the molecule to
be a superposition of charm mesons:
$D^{*0} \bar D^0 + D^0 \bar D^{*0}$.  The interaction of the charm
mesons with pions depends on the pion momentum.
A quantitative analysis of the effects of pions on the $X$
might require all these effects to be taken into account.

Since the new feature of the $X(3872)$ is the near equality between
the mass difference $M_2-M_1$ of the constituents
and the mass $m$ of a meson that can be exchanged between them,
it is worthwhile to analyze this feature in a simple model that
avoids all the other complications of the $X$.
The simplest such model is one
in which the primary constituents $D_1$ and $D_2$ are spin-0 mesons
with a large positive scattering length
and a momentum-independent coupling to a lighter spin-0 meson $\phi$.
We will calculate the effects of the exchanged meson in this model
to next-to-leading order in the $D_2-D_1 \phi$ coupling constant $g$.

In Sec.~\ref{sec:model}, we write down the lagrangian
for the model and discuss its parameters.
In Sec.~\ref{sec:zeroth}, we summarize the results of the model at
0$^{\rm th}$ order in $g$ that were derived
previously in Refs.~\cite{Braaten:2005jj,Braaten:2006sy}.
The results include the  short-distance production rate for $X$
and the line shape of $X$ in a short-distance decay channel.
The ultraviolet divergences can be removed
by renormalization of the coupling constant for the $D_1 D_2$
contact interaction and by renormalization of
short-distance coefficients in the operator product expansion.
In Sec.~\ref{sec:second},  we calculate the observables in
Sec.~\ref{sec:zeroth} to next-to-leading order in $g$.
We also calculate the short-distance production rate
for $D_1 D_1 \phi$ and the decay rate of $X$ into $D_1 D_1 \phi$.
There are new ultraviolet divergences that can be removed
by renormalization of the $D_2$ mass.
We show that after summing a geometric series of next-to-leading
order corrections, all remaining ultraviolet divergences can again be removed
by renormalization of the coupling constant for the $D_1 D_2$
contact interaction and by renormalization of
short-distance coefficients in the operator product expansion.
A further resummation is required to eliminate an infrared
divergence at the $D_1 D_2$ threshold in the next-to-leading
order corrections.
In Sec.~\ref{sec:summary}, we summarize our results.
The two-loop Feynman diagrams
that arise at next-to-leading order in $g$ are calculated in an Appendix.

\section{A Scalar Meson Model}
\label{sec:model}

\subsection{Lagrangian}

\begin{figure}[t]
\includegraphics[width=16cm]{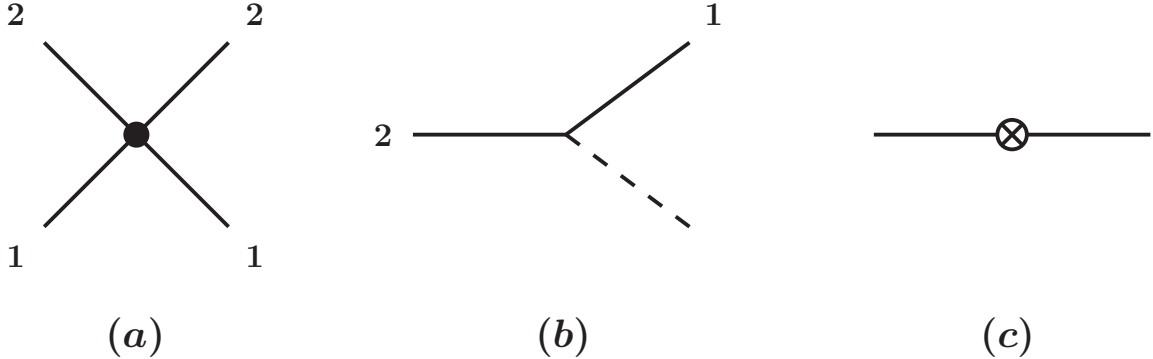}
\caption{
Vertices for (a) the $D_1 D_2$ contact interaction,
(b) the $D_1 D_2 \phi$ interaction,
and (c) the $D_2$ mass counterterm.
\label{fig:vertex}}
\end{figure}

We consider the simplest possible model
with a bound state $X$ whose mass is close to a 3-body threshold.
The primary constituents $D_1$ and $D_2$ of the bound state
are scalar mesons whose masses $M_1$ and $M_2$ satisfy
$M_1< M_2$. There is also a scalar meson $\phi$ whose mass $m$
is close to the mass difference $M_2 - M_1$.
We will refer to this model as the {\it scalar meson model}.
It is defined by a nonrelativistic quantum field theory with three
complex scalar fields $D_1$, $D_2$, and $\phi$.
The free terms in the lagrangian%
\footnote{The superscript $\dagger$ on a field
    indicates its complex conjugate.}
are
\begin{eqnarray}
{\cal L}_{\rm free} 
=\sum_{i=1,\,2}
D_i^\dagger\left(i\frac{\partial}{\partial t}
-M_i+\frac{1}{2M_i}\nabla^2\right)D_i
+\phi^\dagger\left(i\frac{\partial}{\partial t}
-m+\frac{1}{2m}\nabla^2\right)\phi .
\label{Lfree}
\end{eqnarray}
The interaction terms for the scalar meson model are
\begin{equation}
{\cal L}_{\rm int} =
- \lambda_0 D_1^\dagger D_2^\dagger D_1 D_2
- g \big( D_2^\dagger D_1 \phi + D_1^\dagger \phi^\dagger D_2 \big)
- \delta M D_2^\dagger D_2.
\label{Lint}
\end{equation}
The interaction vertices are illustrated in Fig.~\ref{fig:vertex}.
The coupling constants $\lambda_0$
and $g$ have mass dimensions $-2$ and $-1/2$, respectively.
We assume that there is a fine-tuning that makes the
$D_1 D_2$ scattering length large, which implies that the
$D_1 D_2$ contact interaction with coupling constant $\lambda_0$
must be treated nonperturbatively.
The subscript on $\lambda_0$ emphasizes that it is a
bare coupling constant that requires renormalization.
We assume that the $D_2-D_1 \phi$ interaction can be treated
perturbatively and we calculate its effects through order $g^2$.
At this order, no renormalization of $g$ is required.
However, we will see that nonperturbative renormalization of
$\lambda_0$ requires
summing a geometric series to all orders in $g^2$.
There are also ultraviolet divergent
corrections to the mass of the $D_2$ which are cancelled
by the mass counterterm $\delta M$ in Eq.~(\ref{Lint}).

\subsection{Masses}

It is convenient to introduce concise notations for various
combinations of the masses.
The sum and difference of the $D_1$ and $D_2$ masses are
\begin{subequations}
\begin{eqnarray}
M_{1+2} &=&  M_1+M_2 ,
\label{M1plus2}
\\
M_{2-1} &=&  M_2-M_1 .
\label{M2minus}
\end{eqnarray}
\end{subequations}
The reduced masses of $D_1 D_2$
and of $D_1 \phi$ are
\begin{subequations}
\begin{eqnarray}
M_{12} &=& \frac{M_1 M_2}{M_1+M_2},
\\
m_1 &=& \frac{M_1 m}{M_1+m} .
\end{eqnarray}
\label{Mred}
\end{subequations}

If $m$ is much smaller than $M_1$,
the natural energy and 3-momentum scales associated with exchange
of the meson $\phi$ between the mesons $D_1$ and $D_2$
are $m^2/M_{12}$ and $m$, respectively.
We assume that the energy gap between the $D_2$ mass and
the $D_1 \phi$ threshold is small compared to
the natural energy scale:
\begin{equation}
M_{2-1} - m  \ll m^2/M_{12} .
\label{dM-small}
\end{equation}
The total number of heavy mesons $D_1$ and $D_2$ is conserved by the
interactions in Eq.~(\ref{Lint}).
We are interested in the sector of the theory with two heavy mesons
and, more specifically,
in the threshold region where the invariant mass $M$
of all the particles is very close to $M_{1+2}$:
\begin{equation}
|M - M_{1+2}| \ll m^2/M_{12} .
\label{E-small}
\end{equation}
This restricts the possible scattering states
to $D_1 D_2$ and $D_1 D_1 \phi$.

For numerical illustrations, we will use masses
that correspond to the $D^{*0} \bar D^0/D^0 \bar D^0 \pi^0$ system.
The meson masses are $M_2=2006.7$ MeV, $M_1=1864.5$ MeV, and $m = 135.0$ MeV.
Thus $M_{12} \approx 966$ MeV and $M_{2-1} - m = 7.2$ MeV.
The natural energy scale is $m^2/M_{12} \approx 19$ MeV.
The condition in Eq.~(\ref{dM-small}) is only marginally satisfied:
$M_{12} (M_{2-1} - m)/m^2 \approx 0.4$.
We take the binding energy of $X$ to be $E_X = 0.4$ MeV,
which is near the central value of the measurement of $M_X$ in
Eq.~(\ref{MX-CLEO}).  If we express this binding energy as
$E_X = \kappa_X^2/(2 M_{12})$, the binding momentum is
$\kappa_X = 27.8$ MeV.

\subsection{Coupling constants}

We impose an ultraviolet cutoff on the momenta
$\vec p \,$ of particles in the center-of-momentum frame
that restricts them to the region $|\vec p \,| \ll m$.
This guarantees that all the particles remain nonrelativistic.
It was shown in Ref.~\cite{Braaten:2006sy} that at
leading order $g^0$, all ultraviolet divergences
can be absorbed into $\lambda_0$
and into short-distance coefficients in the operator product expansion.
At order $g^2$, there are new ultraviolet
divergences that can be absorbed into the $D_2$ mass.
We will find that the remaining ultraviolet divergences
can again be absorbed into $\lambda_0$
and into short-distance coefficients in the operator product expansion.

The coupling constant $\lambda_0$ must be tuned to near the
critical value at which the $D_1 D_2$ scattering length diverges
in order for the bound state $X$ to have a small binding energy
$E_X \ll m^2/M_{12}$.  In order to compare results at
order $g^0$ and order $g^2$,
we must have a well-defined renormalization prescription for $\lambda_0$
or, equivalently, a clear definition of the binding energy $E_X$.
We will find that all the amplitudes for processes near the
$D_1 D_2$ threshold have a resonant term with a common factor
$\mathcal{A}(E)$ that depends on the energy $E$ in the rest frame.
For example, the line shape in a short-distance decay channel of
$X$ produced by a short-distance process has a long-distance factor
$|\mathcal{A}(M)|^2$, where $M$ is the invariant mass of the decay products.
The square of the resonant factor can be expressed in the form
\begin{equation}
|\mathcal{A}(E)|^2 =
\frac{1}{ [{\rm Re} \, \mathcal{A}^{-1}(E)]^2
      + [{\rm Im} \, \mathcal{A}^{-1}(E)]^2} .
\end{equation}
As a function of the real energy $E$,
this resonance factor has a peak when $E$ is close to $M_{1+2}- E_X$.
The peak arises because ${\rm Re} \, \mathcal{A}^{-1}(E)$ vanishes
near that value of $E$  and ${\rm Im} \, \mathcal{A}^{-1}(E)$ is much smaller
than ${\rm Re} \, \mathcal{A}^{-1}(E)$ except near that value of $E$.
As our renormalization prescription for $\lambda_0$,
we demand that ${\rm Re} \, \mathcal{A}^{-1}(E)$ vanishes
at an energy determined by a real variable $\kappa_X$:
\begin{equation}
{\rm Re} \, \mathcal{A}^{-1}(E) = 0
\hspace{1cm}
{\rm at \ } E = M_{1+2} - \kappa^2_X/(2 M_{12}).
\label{ReAinv:zero}
\end{equation}
The dependence of an observable on the bare parameter $\lambda_0$
can be eliminated in favor of the renormalization parameter $\kappa_X$.
We assume that $\kappa_X$ is small compared to the natural momentum scale:
$\kappa_X \ll m$.

The amplitude $\mathcal{A}(E)$ has a pole at a complex
energy $E_{\rm pole}$ near the real energy at which
$|\mathcal{A}(E)|^2$ has its maximum.
We choose to define the mass $M_X$ and the width
$\Gamma_X$ of $X$ by expressing that complex energy in the form
\begin{eqnarray}
E_{\rm pole} \equiv  M_X - i \Gamma_X/2.
\label{Epole}
\end{eqnarray}
We also choose to define the binding energy $E_X$ by
\begin{eqnarray}
E_X \equiv  M_{1+2} - M_X.
\label{MX-EX}
\end{eqnarray}
We assume that the binding energy
and the width of $X$ are both small
compared to the natural energy scale:
$E_X, \Gamma_X \ll m^2/M_{12}$.

In Ref.~\cite{Braaten:2006sy}, the authors proposed a renormalization
scheme in which $\lambda_0$ was tuned to give a prescribed value
for the complex parameter $E_{\rm pole}$.
The binding energy $E_X$ and the width $\Gamma_X$
were taken as the input parameters that determine $E_{\rm pole}$.
This renormalization scheme requires $\lambda_0$ to be a complex
coupling constant.  The imaginary part of $\lambda_0$ takes into
account inelastic scattering channels for $D_1D_2$ that
are outside the energy range described by the effective field theory.
In this paper, we assume that any such inelastic
channels do not give a significant contribution to the width of $X$.
The coupling constant $\lambda_0$ is therefore real valued.
We could use the binding energy $E_X$ as the input parameter
and eliminate $\lambda_0$ in favor of $E_X$, and then
$\Gamma_X$ would be calculable.
We find it more convenient to eliminate $\lambda_0$ in favor of the
variable $\kappa_X$ introduced in Eq.~(\ref{ReAinv:zero}).
In this renormalization scheme, $E_X$ and $\Gamma_X$ are both calculable
in terms of $\kappa_X$, $g$, and the masses $M_1$, $M_2$, and $m$.

The most convenient input for determining the
coupling constant $g$ is the width $\Gamma_2$
of the heavy meson $D_2$.  Since $M_2 > M_1 + m$,
the $D_1 D_2 \phi$ vertex in Fig.~\ref{fig:vertex}(b)
allows $D_2$ to decay into $D_1 \phi$.
Calculating the width using nonrelativistic phase space,
we obtain
\begin{equation}
\Gamma_2 =
\frac{g^2 m_1}{\pi}
\left[ 2 m_1 (M_{2-1} - m) \right]^{1/2},
\label{Gam2-NR}
\end{equation}
where $M_{12}$ and $m_1$ are the reduced masses defined in
Eqs.~(\ref{Mred}).
The value of the coupling constant $g$
can be determined from $\Gamma_2$ and the masses.
We assume that the width of $D_2$ is small compared to the
natural energy scale: $\Gamma_2 \ll m^2/M_{12}$.
We also assume $\Gamma_2$ is small enough to allow perturbation
theory in the coupling constant $g$.

For numerical illustrations, we will use values of the parameters
that correspond to the $D^{*0} \bar D^0/D^0 \bar D^0 \pi^0$ system.
We take the width of $D_2$ to be $\Gamma_2 = 0.07$ MeV,
which is approximately the total width for $D^{*0}$ \cite{Braaten:2005jj}.
This is much smaller than the natural energy scale $m^2/M_{12}$
and also rather small compared to $M_{2-1}-m$:
$\Gamma_2/(M_{2-1}-m) \approx 0.01$.
Using Eq.~(\ref{Gam2-NR}),
we determine the numerical value of the $D_1 D_2 \phi$
coupling constant to be $g = 0.0064$ MeV$^{-1/2}$.
When considering the dependence of observables on $g$,
it will be more convenient to regard them as functions of $\Gamma_2$.

We now list all the energies that are assumed to be small compared
to the natural scale $m^2/M_{12}$:
\begin{itemize}
\item
the energy $M-M_{1+2}$ of a $D_1 D_2$ or $D_1 D_1 \phi$ system
relative to the $D_1 D_2$ threshold,
\item
the energy difference $M_{2-1} - m$
between the $D_2$ mass and the $D_1 \phi$ threshold,
\item
the decay width $\Gamma_2$ of $D_2$, which is given in Eq.~(\ref{Gam2-NR}),
\item
the binding energy $E_X$ of the molecule,
which is approximately equal to $\kappa_X^2/(2 M_{12})$,
where $\kappa_X$ is a renormalization parameter.
\end{itemize}
We will carry out our calculations under the assumption
that these energies are all comparable and
that the masses $M_2$, $M_1$, and $m$ are also comparable.
The only approximations we will make are those that can be justified
by a nonrelativistic approximation, such as $M_{2-1}-m \ll M_{12}$.

\subsection{Self-energy of $\bm{D_2}$}

\begin{figure}[t]
\includegraphics[width=14cm]{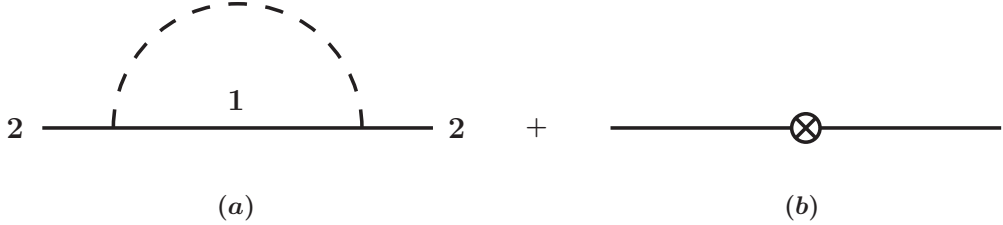}
\caption{
The diagrams for the $D_2$ self-energy.
\label{fig:self}}
\end{figure}

At order $g^2$, renormalization of the $D_2$ mass is necessary.
In the {\it on-shell renormalization scheme}, the parameter $M_2$
in the free lagrangian in Eq.~(\ref{Lfree}) is chosen to be the
physical $D_2$ mass.  The physical mass $M_2$ and the width
$\Gamma_2$ of $D_2$ can be defined by specifying that
the exact $D_2$ propagator at 0 momentum
has a pole in the energy at $p_0 = M_2 - i \Gamma_2/2$.
The pole in the energy $p_0$ at 0 momentum
is the solution to the equation
\begin{equation}
p_0 - M_2 - \Sigma(p_0,0) = 0,
\label{pole-eq}
\end{equation}
where $\Sigma(p_0,p)$ is the $D_2$ self-energy.
The exact $D_2$ self-energy is the sum of the
one-loop diagram in Fig.~\ref{fig:self}(a)
and the mass counterterm in Fig.~\ref{fig:self}(b):
\begin{equation}
\Sigma(p_0,p) = \Sigma_{2a}(p_0,p) + \delta M.
\label{Sigma:exact}
\end{equation}
There are no other diagrams for the self-energy at higher order in $g$.
The one-loop diagram is calculated in Section~\ref{sec:Self-energy}
of the Appendix.  The analytic expression for the self-energy diagram
using dimensional regularization
is given in Eq.~(\ref{Sigma2-dimreg0}).
The result in a general regularization scheme is
\begin{eqnarray}
\Sigma_{2a}(p_0,p) = \Sigma_{2a}(M_1+m,0)
- i \frac{g^2 m_1}{2 \pi}
\left( 2 m_1 \left[ p_0 - M_1 - m - \frac{p^2}{2(M_1+m)} + i
\varepsilon \right] \right)^{1/2} .
\label{D2self}
\end{eqnarray}
The term $\Sigma_{2a}(M_1+m,0)$ is real valued
and includes a linear ultraviolet divergence.
The divergence is cancelled by a divergent term in the mass counterterm
$\delta M$ in Eq.~(\ref{Sigma:exact}). In the on-shell renormalization
scheme, the finite terms in $\delta M$ are chosen so that the
solution to Eq.~(\ref{pole-eq}) satisfies ${\rm Re} \, p_0 = M_2$.
The wavefunction normalization factor for a $D_2$ with momentum $p$ is
\begin{equation}
Z_2(p)^{-1} = 1 -
\frac{\partial \ }{\partial p_0} \Sigma(M_2 - i \Gamma_2/2 + p^2/(2M_2), p) .
\label{Z2-def}
\end{equation}

Since Eq.~(\ref{pole-eq}) cannot be solved analytically,
the on-shell renormalization scheme is not the most convenient choice.
A simpler renormalization scheme for the $D_2$ mass is to choose
the mass counterterm so the self-energy $\Sigma_{2a}(p_0,0)$
vanishes at the $D_1 D_2$ threshold:
$\delta M = - \Sigma_{2a}(M_1+m,0)$.
The self-energy is then
\begin{eqnarray}
\Sigma(p_0,p) =
- i \frac{g^2 m_1}{2 \pi}
\left( 2 m_1 \left[ p_0 - M_1 - m - \frac{p^2}{2(M_1+m)} + i
\varepsilon \right] \right)^{1/2} .
\label{D2self:ren}
\end{eqnarray}
With this renormalization prescription, the parameter $M_2$
differs at order $g^4$ from the physical $D_2$ mass,
which is given by the real part of the solution to Eq.~(\ref{pole-eq}).
The width $\Gamma_2$, which is given by the imaginary part of the solution
to Eq.~(\ref{pole-eq}), reduces at leading order in $g$ to
$\Gamma_2 = - 2 \, {\rm Im} \, \Sigma_{2a}(M_2,0)$.
This agrees with the explicit expression for the decay rate of $D_2$
at order $g^2$ obtained in Eq.~(\ref{Gam2-NR}).
To order $g^2$, the wavefunction normalization factor
in Eq.~(\ref{Z2-def}) is
\begin{equation}
Z_2(p)^{-1} = 1 + i
\frac{g^2 m_1^2}{2 \pi}
\Big( 2 m_1 (M_{2-1} - m)[1 - p^2/(2(M_1+m)M_2)] \Big)^{-1/2} .
\end{equation}
Since the momentum $p$ is restricted to the nonrelativistic region,
the term $p^2/(2(M_1+m)M_2)$ can always be neglected compared to 1.
The resulting
wavefunction normalization factor is independent of $p$:
\begin{equation}
Z_2^{-1} = 1 + i
\frac{g^2 m_1^2}{2 \pi} \left[ 2 m_1 (M_{2-1} - m) \right]^{-1/2}.
\label{Z2-NR}
\end{equation}
The correction term is pure imaginary.

\subsection{The Complex Mass Scheme}

The $D_2-D_1 \phi$ interaction allows a virtual $D_2$
whose energy is greater than $M_1 + m$ to decay into $D_1\phi$.
In the amplitude for such a process at leading order in $g$,
the propagator of the virtual $D_2$ diverges.
Near such a point in momentum space, the corrections to the
$D_2$ propagator from higher orders in $g$ are not suppressed.
This problem can be solved by summing the
$D_2$ propagator corrections to all orders.
The effect is to add $-\Sigma(p_0,p)$ to the denominator
$p_0 - p^2/(2M_2) + i \varepsilon$ of the $D_2$ propagator.
The imaginary part of the self-energy $\Sigma(p_0,p)$
eliminates any divergence in the $D_2$ propagator
for real values of the energy.  One drawback of this
solution is that it leads to unnecessarily complicated expressions
for the integrals in loop diagrams with $D_2$ propagators.

Summing the $D_2$ self-energy corrections to all orders
is essential only near the pole in the propagator.
In other regions of the energy $p_0$, $\Sigma(p_0,p)$
is a perturbative correction to the denominator of the propagator
that is suppressed by a power of $g^2$.  In those regions,
summing the $D_2$ propagator corrections to all orders is optional.
At the pole in $p_0$, the self-energy $\Sigma(p_0,p)$
reduces to $-i \Gamma_2/2$.
An alternative resummation that eliminates the divergence in the
propagator for real values of the energy without changing the
order in $g$ of the truncation error is to sum only the $-i \Gamma_2/2$ term
in the self-energy to all orders and treat the remainder
of the self-energy as a perturbation.  An advantage of this partial
resummation is that it leads to simpler expressions for the integrals
in loop diagrams with $D_2$ propagators.

A systematic method for implementing this partial resummation
is the {\it complex mass scheme} \cite{Denner:2006ic}.
This method has been used in Standard Model calculations at
next-to-leading order
\cite{Denner:2005es,Denner:2005fg,Bredenstein:2006rh}.
The application of this method to the scalar meson model
involves adding cancelling terms to the free
and interaction terms in the Lagrangian:
\begin{subequations}
\begin{eqnarray}
\Delta {\cal L}_{\rm free} &=&
(i \Gamma_2/2) D_2^\dagger D_2
+ (Z_2^{-1}-1) D_2^\dagger \left( i\frac{\partial}{\partial t}
- M_2 + i \Gamma_2/2 + \frac{1}{2M_2} \nabla^2 \right) D_2,
\label{DLfree}
\\
\Delta {\cal L}_{\rm int} &=&
- (i \Gamma_2/2) D_2^\dagger D_2
- (Z_2^{-1}-1) D_2^\dagger \left( i\frac{\partial}{\partial t}
- M_2 + i \Gamma_2/2 + \frac{1}{2M_2} \nabla^2 \right) D_2 ,
\label{DLint}
\end{eqnarray}
\end{subequations}
where $\Gamma_2$ is the $D_2$ width and $Z_2$ is the $D_2$
wavefunction normalization factor.
To order $g^2$, $\Gamma_2$ and $Z_2$ are given by
Eqs.~(\ref{Gam2-NR}) and (\ref{Z2-NR}).
Since $i \Gamma_2/2$ is pure imaginary and $Z_2$
is complex, the free Lagrangian consisting of the sum of
Eqs.~(\ref{Lfree}) and (\ref{DLfree}) corresponds to a nonhermitian
Hamiltonian.  The effects of the nonhermiticity are
cancelled exactly if the interaction terms in Eq.~(\ref{DLint})
are calculated to all orders.
At any finite order in perturbation theory, the complex mass scheme
gives amplitudes that are uniformly accurate to the appropriate
order in $g$, even if there is a virtual $D_2$ that decays into $D_1 \phi$.
The effect of adding Eq.~(\ref{DLfree}) to the free Lagrangian in
Eq.~(\ref{Lfree}) is to change the $D_2$ propagator:
\begin{equation}
\frac{i}{p_0 - M_2 - p^2/(2M_2) + i \varepsilon} \longrightarrow
\frac{i Z_2}{p_0 - (M_2 - i\Gamma_2/2) - p^2/(2M_2)} .
\label{newpropagator}
\end{equation}
The effect of adding Eq.~(\ref{DLint}) to the interaction Lagrangian in
Eq.~(\ref{Lint}) is to change the counterterm vertex in
Fig.~\ref{fig:vertex}:
\begin{equation}
-i \delta M \longrightarrow
-i ( \delta M + i\Gamma_2/2 )
- i (Z_2^{-1} - 1) \left[ p_0 - M_2 + i\Gamma_2/2 - p^2/(2M_2) \right] .
\label{newcounterterm}
\end{equation}
It is convenient to also rescale the fields $D_2(x)$ and
$D_2^\dagger(x)$ by the same complex factor $Z_2^{1/2}$.
This eliminates the factor of $Z_2$ from the numerator of the
$D_2$ propagator in Eq.~(\ref{newpropagator}).
It also multiplies all the interaction terms in Eqs.~(\ref{Lint})
and (\ref{DLint}) by $Z_2$ or $Z_2^{1/2}$.
Since we will use nonperturbative renormalization for the coupling
constant $Z_2 \lambda_0$ for the $D_1 D_2$ contact interaction,
the factor of $Z_2$ only affects the value of the unphysical
bare coupling constant $\lambda_0$.
Since $Z_2 = 1 + O(g^2)$,
the other factors of $Z_2$ or $Z_2^{1/2}$ in the interaction vertices
contribute only at order $g^4$ and higher.
We will be calculating only to order $g^2$, so
we will ignore these factors of $Z_2$ and $Z_2^{1/2}$.

The perturbation expansion in the complex mass scheme corresponds
to ordinary perturbation theory in $g$ followed by a resummation
to all orders of a subset of terms that are higher order in $g$.
We will refer to calculations to order $g^0$ in this
perturbation expansion as {\it leading order (LO)} in the complex mass scheme.
We will refer to calculations through order $g^2$ in this
perturbation expansion as {\it next-to-leading order (NLO)}
in the complex mass scheme.

\subsection{Propagator for $\bm{X}$}
\label{sec:PropX}

We can use the scalar meson model to describe processes with
asymptotic states that include not only the particles $D_1$
and $\phi$, but also the bound state $X$. The local composite
operator $D_1^\dagger D_2^\dagger(x)$ has a nonzero amplitude to
create $X$ from the vacuum. Thus $D_1 D_2(x)$ can be used as an
interpolating field for $X$. It is more convenient to use the
operator $\lambda_0 D_1 D_2(x)$, because this is a renormalized
operator whose matrix elements do not depend on the ultraviolet
cutoff \cite{Braaten:2006sy}.
The corresponding propagator for $X$ is
\begin{equation}
\Delta_X(E,P) =
\int d^4x \, e^{i P\cdot x}
\langle \emptyset | \lambda_0 D_1 D_2(x) \lambda_0 D_1^\dagger D_2 ^\dagger(0)
    | \emptyset \rangle.
\label{X-prop}
\end{equation}
The momentum 4-vector in the Fourier transform is
$P^\mu = (E,\vec P \,)$, where $E$ and $\vec P \,$ are the energy
and momentum of $X$.
At $P=0$, this propagator
has a pole at the complex energy $E_{\rm pole}$.
Near the pole in the energy, the behavior of the propagator
at $P=0$ is
\begin{equation}
\Delta_X(E,0) \longrightarrow
\frac{i Z_X}{E - E_{\rm pole}+i\varepsilon} .
\label{DeltaX-pole}
\end{equation}
The energy $E_{\rm pole}$ determines the binding energy $E_X$
and the width $\Gamma_X$ of $X$ through Eq.~(\ref{Epole}).
The residue of the pole defines the wavefunction normalization
factor $Z_X$, which is complex valued.
Because the propagator $\Delta_X(E,P)$
has mass dimension $-2$, $Z_X$ has mass dimension $-1$.

Strictly speaking, since $X$ has a nonzero width $\Gamma_X$,
there are no T-matrix elements for processes involving $X$
in the initial or final state,
because the decay of the $X$ prevents it from
being a truly asymptotic state.  However if $\Gamma_X$ is
sufficiently small, the $X$ can propagate over a long
time interval before decaying, and it can therefore be treated as a
quasi-asymptotic state.
We can use the Lehmann-Symanzik-Zimmermann
(LSZ) formalism to define T-matrix elements for
the $X$ that are closely related to T-matrix elements for
decay products of the $X$ whose total energy is tuned
to the peak of the $X$ resonance.
To define the T-matrix element ${\cal T}$ for
a process with $X$ in the initial or final state, we start
with a Green's functions for the operator
$\lambda_0 D_1^\dagger D_2^\dagger(x)$ or $\lambda_0 D_1D_2(x)$.
The connected Green's function at 4-momentum $(E,\vec P \,)$
is amputated by multiplying by the inverse propagator
$\Delta_X(E,P)^{-1}$,
it is normalized by multiplying by $Z_X^{1/2}$,
and then it is evaluated at the real energy
$M_X + P^2/(2 M_X)$. This gives the T-matrix element multiplied by $i$
for a state $X$ with the standard nonrelativistic normalization.
The T-matrix element ${\cal T}$ for a state $X$ with the standard relativistic
normalization is obtained by multiplying by $\sqrt{2 M_X}$.

To justify this prescription, we consider a process $A \to B + C$,
where $A$ and $B$ both represent one or more particles and $C$
denotes a set of particles that can be decay products of $X$.
The T-matrix for this process will have a resonant enhancement
when the invariant mass $M$ of the particles in $C$ is near the
mass $M_X$.  The resonant contribution to the T-matrix element
for $A \to B + C$ in the rest frame of $C$ can be approximated by
\begin{equation}
i {\cal T}[A \to B + C] \approx i {\cal T}[A \to B + X] \,
\frac{i}{2 M_X(M - M_X + i \Gamma_X/2)}  \, i {\cal T}[X \to C].
\label{TABC:res}
\end{equation}
Our prescription for T-matrix elements such as ${\cal T}[A \to B + X]$
and ${\cal T}[X \to C]$ involves evaluating the amputated connected
Green's function at the real energy $E = M_X$.  This ensures that
the expression in Eq.~(\ref{TABC:res}) is as good an approximation
as possible to the T-matrix element ${\cal T}[A \to B + C]$
near the peak of the resonance.

As a simple illustration of the prescription for T-matrix elements
for processes involving $X$, we consider the forward scattering
process $X \to X$.  The relevant Green's function in the rest
frame of $X$ is the negative of the propagator $\Delta_X(E,0)$.  This must
be multiplied by two factors of $\Delta_X(E,0)^{-1} Z_X^{1/2}$,
one for the $X$ in the inital state and one for the $X$
in the final state.  The T-matrix element $i{\cal T}[X \to X]$
is then obtained by evaluating this at $E=M_X$
and multiplying it by $2 M_X$:
\begin{equation}
i{\cal T}[X \to X] = - \frac{2 M_X Z_X}{\Delta_X(M_X,0)} .
\label{TXX}
\end{equation}

\subsection{Short-distance production and decay processes}

The scalar meson model defined by the lagrangian
in Eqs.~(\ref{Lfree}) and (\ref{Lint}) can be a low-energy
approximation to a more fundamental quantum field theory.
The fundamental theory may include
high energy processes that can create $D_1 D_2$ or $D_1 D_1 \phi$
with invariant mass near $M_{1+2}$.
If {\it long-distance} effects involving momenta
much smaller than $m$ can be separated from {\it short-distance}
effects involving momenta of order $m$ and larger,
we can use the scalar meson model
to calculate the long-distance effects.
The basic tool required to separate long-distance effects from
short-distance effects is the {\it operator product expansion}
\cite{Braaten:2006sy}.
Applications of the operator product expansion to the scalar meson model
at 0$^{\rm th}$ order in $g$ were described
in Ref.~\cite{Braaten:2006sy}.
Below, we give the leading terms in the operator product expansions
of the T-matrix elements for some of the processes considered
in Ref.~\cite{Braaten:2006sy}.

\begin{figure}[t]
\includegraphics[width=2.5cm]{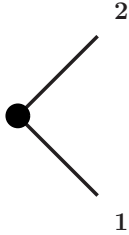}
\caption{
Vertex for the $\lambda_0 D_1 D_2$ and
$\lambda_0 D_1^\dagger D_2^\dagger$ operators.
\label{fig:operators}}
\end{figure}

The general production process for $X$ has the form
$A \to B + X$, where $A$ and $B$ each represent one or more particles.
There can be analogous production processes for
$D_1 D_1 \phi$.  In the case of $D_1 D_1 \phi$,
we assume that the invariant mass $M$ satisfies the
condition in Eq.~(\ref{E-small}).  This implies that the relative
momenta of the mesons, which we denote generically by $\vec p\,$,
are small compared to the natural momentum scale $m$.
We call the production process
a short-distance process if all the particles in $A$ and
$B$ have momenta in the rest frame of $X$ or
$D_1 D_1 \phi$ that are of order $m$ or
larger.  The T-matrix element for a short-distance production
process can be expanded in powers of the small energy
differences $M-M_{1+2}$ and $M_{2-1} - m$ divided by $m^2/M_{12}$
and higher energy scales and in powers of the small relative momenta
$\vec p \,$ divided by $m$ and higher momentum scales.
The operator product expansion can be used to
organize the expansions of the T-matrix elements into
sums of products of short-distance coefficients and matrix elements
of local operators between the vacuum state $| \emptyset \rangle$
and the appropriate final state.
The leading terms in the expansions are those with the
lowest dimension operator, which is $D_1^\dagger D_2^\dagger(x)$.
The corresponding renormalized operator is
$\lambda_0 D_1^\dagger D_2^\dagger(x)$.
In Feynman diagrams, the local
operator $\lambda_0 D_1^\dagger D_2^\dagger(x)$ is represented by a dot
from which a $D_1$ line and a $D_2$ line emerge, as illustrated in
Fig.~\ref{fig:operators}. If we keep only the leading terms, the
factorization formulas for the T-matrix elements reduce to%
\footnote{The Wilson coefficients ${\cal C}_A^{B,12}$ and
${\cal C}_{12}^{C}$ differ from those
in Ref.~\cite{Braaten:2006sy} by a factor of $\lambda_0$.
This choice simplifies many equations.}\setcounter{footnote}{1}
\begin{subequations}
\begin{eqnarray}
{\cal T}[A \to B + X]  &=& \sqrt{2M_X} \; {\cal C}_A^{B,12} \,
\langle X| \lambda_0 D_1^\dagger D_2^\dagger(0)
    | \emptyset \rangle,
\label{TAXpB}
\\
{\cal T}[A \to B + D_1 D_1 \phi] &=& \sqrt{8M_1^2m} \;
{\cal C}_A^{B,12} \, \langle D_1 D_1 \phi |
    \lambda_0 D_1^\dagger D_2^\dagger(0) | \emptyset \rangle.
\label{TADDphiB0}
\end{eqnarray}
\label{TADDX}
\end{subequations}
The Wilson coefficient ${\cal C}_A^{B,12}$ is
a function of high energy scales, such as $m$ and $M_{12}$, and of
the total momentum of the $X$ or $D_1 D_1 \phi$. The
only dependence on whether the final state contains $X$
or $D_1 D_1 \phi$ is in the matrix elements of the local operator
$\lambda_0 D_1^\dagger D_2^\dagger(0)$.
Short-distance and long-distance effects
are separated in Eqs.~(\ref{TADDX}).

The decay of $X$ into $D_1 D_1 \phi$
can be described within the scalar meson model.
The fundamental theory may also allow the decay of $X$
into other final states $C$.
We call such a process a short-distance decay
if all the particles in $C$ have momenta
in the rest frame that are of order $m$ or larger.
The T-matrix element for such a
process can be expanded in powers of the small energy
differences $M_X-M_{1+2}$ and $M_{2-1} - m$ divided by $m^2/M_{12}$
and higher energy scales.  The operator product expansion
can be used to organize the expansion of the T-matrix element
into sums of products of short-distance coefficients and matrix elements
of local operators between the state $| X \rangle$ and the vacuum state.
  The leading term
in the expansion is the one with the lowest dimension operator,
which is $D_1 D_2(x)$.  The corresponding renormalized operator is
$\lambda_0 D_1 D_2(x)$.  If we keep only this term, the
factorization formula for the T-matrix element reduces to\footnotemark
\begin{equation}
{\cal T}[X \to C]  = \sqrt{2M_X} \; {\cal C}_{12}^{C} \, \langle
\emptyset | \lambda_0 D_1 D_2(0) | X \rangle.
\label{TXCM}
\end{equation}
The Wilson coefficient ${\cal C}_{12}^{C}$ is
a function of high energy scales, such as $m$ and $M_{12}$.
Short-distance and long-distance effects
are separated in Eqs.~(\ref{TXCM}).

If the fundamental theory includes processes
that allow the production of $X$ via $A \to B + X$
and the decay of $X$ via $X \to C$, it also allows the process
$A \to B + C$, where $C$ represents the same particles
but with a variable invariant mass $M$ instead of $M_X$.
This process has a resonant enhancement when $M$
is near the $D_1 D_2$ threshold as specified by Eq.~(\ref{E-small}).
We call it a {\it short-distance} process if each of the particles
in $A$ and $B$ has momentum large compared to $m$ in the rest frame
of $C$ and if the relative momentum between each particle in $C$
and each particle in $A$ or $B$ is large compared to $m$.
The T-matrix element for such a process
can be described within the scalar meson model by a double
operator product expansion.  The leading terms in this
expansion are
\begin{eqnarray}
{\cal T}[A \to B + C]  &=& {\cal C}_A^{B,\,C} + {\cal C}_A^{B,12}
{\cal C}_{12}^C \int d^4x \, e^{i P \cdot x} \langle \emptyset |
\lambda_0 D_1 D_2(x) \lambda_0 D_1^\dagger D_2^\dagger(0) |
\emptyset \rangle,
\label{T:ABCM}
\end{eqnarray}
where the 4-vector is $P^\mu = (M,\vec 0 \,)$. The
Wilson coefficients ${\cal C}_A^{B,12}$ and ${\cal C}_{12}^C$ are
the same ones that appear in the operator product expansions in
Eqs.~(\ref{TADDX}) and (\ref{TXCM}). The first term ${\cal C}_A^{B,C}$
on the right side of Eq.~(\ref{T:ABCM}) takes into account the direct
production of $C$ at short distances.  This term can be expanded
in powers of the small energy differences $M-M_{1+2}$
and $M_{2-1}-m$ divided by energy scales of order $m^2/M_{12}$
or higher. The leading term in this expansion is simply a constant.
According to Eq.~(\ref{X-prop}), the Fourier transform in
Eq.~(\ref{T:ABCM}) is just the $X$ propagator evaluated at $E=M$.
Short-distance and long-distance effects
are not yet separated in Eq.~(\ref{T:ABCM}), because the $X$ propagator
requires an additive renormalization \cite{Braaten:2006sy}.

\section{Leading order in $\bm{g}$}
\label{sec:zeroth}

At $0^{\rm th}$ order in $g$, the only interaction
in the scalar meson model is the contact interaction
between $D_1$ and $D_2$ with coupling constant $\lambda_0$.
The $D_1 D_1 \phi$ states remain
noninteracting at this order in $g$.
In this section, we summarize some of the results at $0^{\rm th}$ order
in $g$ that were obtained in Ref.~\cite{Braaten:2006sy}
and we generalize those results to the complex mass scheme.

\subsection{Amplitude for $\bm{D_1 D_2 \to D_1 D_2}$}

\begin{figure}[t]
\includegraphics[width=5cm]{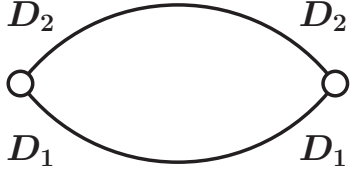}
\caption{ Diagram of order $g^0$ for the amplitude for the
propagation of $D_1D_2$ between contact interactions.
The open dots indicate that the vertex factors $- i \lambda_0$
are omitted.
\label{fig:L-LO}}
\end{figure}

At 0$^{\rm th}$ order in $g$, all the observables for processes
near the $D_1 D_2$ threshold are related in a simple way to the
connected Green's function for $D_1 D_2 \to D_1 D_2$.
The basic building block for this Green's function is the
amplitude $i L_0(E)$ for the propagation of
the $D_1 D_2$ pair between successive contact interactions,
which is given by the Feynman diagram in Fig.~\ref{fig:L-LO}.
The amplitude $L_0(E)$ is calculated in Section~\ref{sec:Propcon0}
of the Appendix.  The analytic expression using dimensional
regularization is given in Eq.~(\ref{L0-3d}).
The result in a general regularization scheme is
\begin{equation}
L_0(E) =  L_0(M_{1+2}) + \frac{M_{12}}{2 \pi} \kappa(E),
\label{L0-sub}
\end{equation}
where $\kappa(E)$ is the energy variable defined by
\begin{equation}
\kappa(E)  =
\sqrt{- 2 M_{12} (E - M_{1+2}) - i \varepsilon} .
\label{kappa-E}
\end{equation}
This variable vanishes at the $D_1D_2$ threshold $E = M_{1+2}$.
It is real and positive if $E < M_{1+2}$
and it is pure imaginary with a negative imaginary part
if $E > M_{1+2}$.
The term $L_0(M_{1+2})$ in Eq.~(\ref{L0-sub}) is real valued and
includes a linear ultraviolet divergence.

\begin{figure}[t]
\includegraphics[width=16cm]{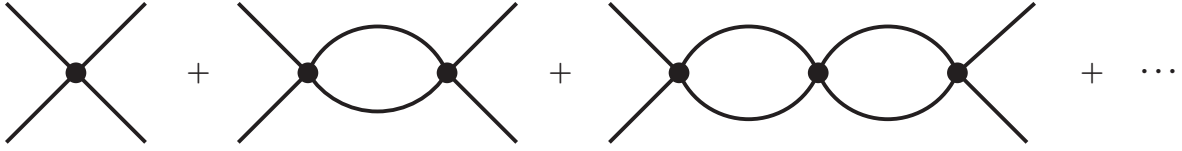}
\caption{
The connected Green's function $i\mathcal{A}_0(E)$
for $D_1 D_2 \to D_1 D_2$ at $0^{\textrm{th}}$ order in $g$
can be obtained by summing a geometric series of one-loop diagrams.
\label{fig:ALO}}
\end{figure}

The amputated connected Green's function $i \mathcal{A}_0(E)$
for $D_1 D_2 \to D_1 D_2$ can be calculated nonperturbatively
by summing the geometric series
represented by Fig.~\ref{fig:ALO} to all orders in $\lambda_0$:
\begin{equation}
i\mathcal{A}_0(E) =
\frac{- i}{1/\lambda_0 -  L_0(E)} .
\label{A-bare}
\end{equation}
The renormalization prescription for $\lambda_0$ specified by
Eq.~(\ref{ReAinv:zero}) is that ${\rm Re}\,\mathcal{A}_0^{-1}(E)$
must vanish at $M_{1+2} - \kappa_X^2/(2M_{12})$.
This requires the inverse bare coupling constant to be
\begin{equation}
\frac{1}{\lambda_0} = L_0(M_{1+2}) +
\frac{M_{12}}{2\pi} \kappa_X .
\label{lam0inv:0}
\end{equation}
Using this expression for $1/\lambda_0$ and
the expression for $L_0(E)$ in Eq.~(\ref{L0-3d}),
the amplitude in Eq.~(\ref{A-bare}) reduces to
\begin{equation}
\mathcal{A}_0(E)  =
\frac{2 \pi/M_{12}}{\kappa(E) - \kappa_X} .
\label{A0-ren}
\end{equation}

If we were to use the complex mass scheme, the amplitude $L_0(E)$
would be given by Eq.~(\ref{L0-sub}) with $M_{1+2}$ replaced by
$M_{1+2} - i \Gamma_2/2$.
If we again demand that ${\rm Re}\,\mathcal{A}_0^{-1}(E)$
must vanish at $M_{1+2} - \kappa_X^2/(2M_{12})$,
the renormalized expression for the amplitude would be
\begin{equation}
\mathcal{A}_0(E)  =
\frac{2 \pi/M_{12}}{\gamma(E) - {\rm Re} \, \gamma_X} ,
\label{A0-cms}
\end{equation}
where $\gamma(E)$ is a complex energy variable defined by
\begin{equation}
\gamma(E)  =
\sqrt{- 2 M_{12} (E - M_{1+2}+ i \Gamma_2/2)}
\label{gamma-E}
\end{equation}
and $\gamma_X$ is its value at $E = M_{1+2} - \kappa_X^2/(2 M_{12})$:
\begin{equation}
\gamma_X  =
\sqrt{\kappa_X^2 - i M_{12} \Gamma_2} .
\label{gammaX}
\end{equation}
The variable $\gamma(E)$ vanishes at the complex threshold
$E = M_{1+2} - i \Gamma/2$ and it has a negative imaginary part
for all real values of $E$.

\subsection{Propagator for $\bm{X}$}
\label{sec:Xprop0}

\begin{figure}[t]
\includegraphics[width=16cm]{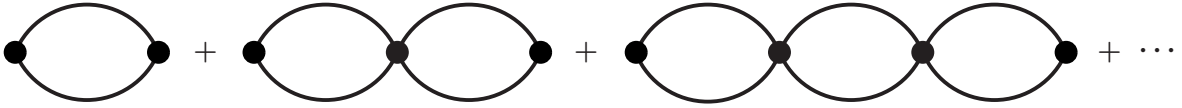}
\caption{
Feynman diagrams for the $X$ propagator at 0$^{\rm th}$ order in $g$.
The interpolating field for the $X$ is $\lambda_0 D_1 D_2$.
\label{fig:Xprop}}
\end{figure}

If the local composite operator
$\lambda_0 D_1 D_2(x)$ is used as an interpolating field for $X$,
the propagator for $X$ is given in Eq.~(\ref{X-prop}).
The diagrams for the propagator of $X$
at 0$^{\rm th}$ order in $g$ are shown in Fig.~\ref{fig:Xprop}.
These diagrams form a geometric series
whose sum in the rest frame $\vec P=0$ is
\begin{equation}
\Delta_X(E,0) =
\frac{i \lambda_0^2 L_0(E)}{1 - \lambda_0 L_0(E)}.
\label{DeltaX-0}
\end{equation}
Using the expression for $\mathcal{A}_0(E)$ in Eq.~(\ref{A-bare}),
we can express the propagator for $X$ as
\begin{equation}
\Delta_X(E,0) = -i [\lambda_0 + \mathcal{A}_0(E)] .
\label{DeltaX:ren}
\end{equation}
To obtain a renormalized propagator for $X$  that does not depend
on the ultraviolet cutoff, an additive renormalization is necessary.
After adding the constant $i \lambda_0$, we obtain the
renormalized propagator $-i \mathcal{A}_0(E)$.

The term  $-i\mathcal{A}_0(E)$ in the propagator for $X$ in
Eq.~(\ref{DeltaX:ren}) has a pole at the complex energy $E_{\rm pole}$
given in Eq.~(\ref{Epole}).
Using the renormalized expression for $\mathcal{A}_0(E)$ in
Eq.~(\ref{A0-ren}), we find that the binding energy and width
of $X$ at 0$^{\rm th}$ order in $g$ are
\begin{subequations}
\begin{eqnarray}
E_X &=& \kappa_X^2/(2M_{12}),
\label{EX:00}
\\
\Gamma_X &=& 0.
\label{GamX:00}
\end{eqnarray}
\label{EGamX:00}
\end{subequations}
The wavefunction normalization factor for $X$ is
\begin{equation}
Z_X = \frac{2 \pi}{M^2_{12}} \kappa_X.
\label{ZX}
\end{equation}

If we were to use the complex mass scheme, the amplitude
$\mathcal{A}_0(E)$ would be given by Eq.~(\ref{A0-cms}).
This amplitude has a pole at a complex energy $E_{\rm pole}$
that determines the binding energy and width of $X$.
The complex parameter $E_{\rm pole}$ satisfies
\begin{equation}
\gamma(E_{\rm pole}) - {\rm Re} \, \gamma_X =0,
\label{gamma-pole:LO}
\end{equation}
where $\gamma(E)$ and $\gamma_X$ are defined in
Eqs.~(\ref{gamma-E}) and (\ref{gammaX}).
Comparing with Eq.~(\ref{Epole}), we find that at
LO in the complex mass scheme, the binding energy and width of $X$ are
\begin{subequations}
\begin{eqnarray}
E_X &=&
( {\rm Re} \, \gamma_X )^2/(2 M_{12}) ,
\label{EX:0}
\\
\Gamma_X &=& \Gamma_2 .
\label{GamX:0}
\end{eqnarray}
\label{EGamX:0}
\end{subequations}
If we treat $\Gamma_2$ as being of order $g^2$, the binding energy
of $X$ in Eq.~(\ref{EX:0}) differs from $\kappa_X^2/(2 M_{12})$
only at order $g^4$.
The wavefunction normalization constant for $X$ at
LO in the complex mass scheme is
\begin{equation}
Z_X = \frac{2 \pi}{M^2_{12}} \,
{\rm Re} \, \gamma_X .
\label{ZX:cms}
\end{equation}

\subsection{Short-distance Production of $\bm{X}$}
\label{sec:Xprod0}

The operator product expansion of the T-matrix element
for a short-distance production process
$A \to B + X$ is given in Eq.~(\ref{TAXpB}).
The vacuum--to--$X$ matrix element can be calculated via the
LSZ formalism using $\lambda_0 D_1D_2(x)$ as an interpolating field
for $X$ \cite{Braaten:2006sy}:
\begin{equation}
\langle X | \lambda_0 D_1^\dagger D_2^\dagger(0) | \,\emptyset \rangle
= Z_X^{1/2} .
\label{XDD0}
\end{equation}
At 0$^{\rm th}$ order in $g$, the normalization constant $Z_X$
is given in Eq.~(\ref{ZX}).  At LO in the
complex mass scheme, $Z_X$ is given in Eq.~(\ref{ZX:cms}).
The T-matrix element in Eq.~(\ref{TAXpB})
can be expressed as the product
of a short-distance factor and a long-distance factor:
\begin{equation}
{\cal T}[A \to B + X]
= \sqrt{2M_X} \, {\cal C}_A^{B,12} \, Z_X^{1/2}.
\label{TAXB-fact}
\end{equation}

The rate for producing $X$ is obtained by squaring the T-matrix element
in Eq.~(\ref{TAXB-fact}) and integrating over the appropriate
phase space.  If $A$ consists of a single particle,
its decay rate into $B + X$ can be expressed
in the factored form
\begin{equation}
\Gamma[A \to B + X] =
\Gamma_A^B \, M_{12} |Z_X|.
\label{GamABX}
\end{equation}
We have followed Ref.~\cite{Braaten:2006sy} in choosing the
long-distance factor in Eq.~(\ref{GamABX}) to be $M_{12} |Z_X|$,
which is dimensionless.
The short-distance factor $\Gamma_A^B$ is
\begin{eqnarray}
\Gamma_A^B &=& \frac{M_{1+2}}{M_A M_{12}} \int \frac{d^3P}{(2
\pi)^3 2E} \int \prod_{i \in B} \frac{d^3p_i}{(2 \pi)^3 2E_i}
\left| {\cal C}_A^{B,12} \right|^2
(2 \pi)^4 \delta^{(4)}(P_A - P - \mbox{$\sum\limits_{i \in B}$}\,p_i) ,
\label{GamADDB}
\end{eqnarray}
where the 4-vector $P^\mu$ satisfies $P^2 = M_{1+2}^2$.

\subsection{Short-distance Decay of $\bm{X}$}
\label{sec:Xdecay0}

The operator product expansion of the T-matrix element
for a short-distance decay process
$X \to C$ is given in Eq.~(\ref{TXCM}).
The $X$--to--vacuum matrix element
can be calculated via the LSZ formalism
using $\lambda_0 D_1^\dagger D_2^\dagger(x)$ as an interpolating field
for $X$ \cite{Braaten:2006sy}:%
\footnote{Our notation might suggest that the matrix element in
Eq.~(\ref{0DDX}) is the complex conjugate of the
matrix element in Eq.~(\ref{XDD0}).  However $| X \rangle$
in Eq.~(\ref{0DDX}) is an in state, while $\langle X |$
in Eq.~(\ref{XDD0}) is the hermitian conjugate of an out state.
These two states are related by the S-matrix:
$| X,\ {\rm out} \rangle = {\cal S} | X,\ {\rm in} \rangle$.}
\begin{equation}
\langle \emptyset| \lambda_0 D_1 D_2(0) | X \rangle
= Z_X^{1/2} .
\label{0DDX}
\end{equation}
At 0$^{\rm th}$ order in $g$, the normalization constant $Z_X$
is given in Eq.~(\ref{ZX}).  At LO in the
complex mass scheme, $Z_X$ is given by Eq.~(\ref{ZX:cms}).
The T-matrix element in Eq.~(\ref{TXCM}) can be expressed as the
product of a short-distance factor and a long-distance factor:
\begin{equation}
{\cal T}[X \to C] =
\sqrt{2M_X} \, {\cal C}_{12}^{C} \, Z_X^{1/2} .
\label{TXCM-fact}
\end{equation}

The decay rate of $X$ into the particles represented by $C$
is obtained by squaring the T-matrix element in Eq.~(\ref{TXCM-fact})
and integrating over the phase space of those particles.
It can be expressed in the factored form
\begin{equation}
\Gamma[X \to C]  =
\Gamma^C \, M_{12} | Z_X |.
\label{GamXCM-fact}
\end{equation}
We have followed Ref.~\cite{Braaten:2006sy} in choosing the
long-distance factor to be $M_{12} | Z_X |$, which is dimensionless.
The short-distance factor $\Gamma^C$ is
\begin{equation}
\Gamma^C = \frac{1}{M_{12}}
\int \prod_{j \in C} \frac{d^3p_j}{(2 \pi)^3 2E_j}
\left| {\cal C}^C_{12} \right|^2
(2 \pi)^4 \delta^{(4)} \!( P - \mbox{$\sum\limits_{j \in C}$} p_j) .
\label{Gam-C}
\end{equation}
where the 4-vector $P^\mu$ satisfies $P^2 = M_{1+2}^2$.

\subsection{Line shape of $\bm{X}$ in a Short-distance Decay Channel}
\label{sec:Xlineshape0}

If $X$ can be produced via the short-distance process $A \to B + X$
and if it can decay  via the short-distance process $X \to C$,
then there can be resonant enhancement of the process
$A \to B + C$, where $C$ represents the same particles
but with a variable invariant mass $M$ instead of $M_X$.
The T-matrix element for this process
can be expressed as the double
operator product expansion in Eq.~(\ref{T:ABCM}).
The Fourier transform of the matrix element is just the $X$ propagator
evaluated at $(E, \vec P \,) = (M, \vec 0 \,)$.
The T-matrix element therefore reduces to
\begin{eqnarray}
{\cal T}[A \to B + C]  &=& {\cal C}_A^{B,C} +
{\cal C}_A^{B,12} {\cal C}_{12}^C
\frac{i \lambda_0^2 L_0(M)}{1-\lambda_0 L_0(M)}.
\label{T:ABC-bare}
\end{eqnarray}
The T-matrix element can be expressed in a form in which the
short-distance effects and long-distance effects are separated:
\begin{eqnarray}
{\cal T}[A \to B + C]  &=& -i{\cal C}_A^{B,12} {\cal C}_{12}^C
 \left[ \mathcal{A}_0(M) - (2 \pi/M_{12}) c_A^{B,C}
\right] ,
\label{T:ABC-simple}
\end{eqnarray}
where $\mathcal{A}_0(E)$ is given in Eq.~(\ref{A0-ren}) and $c_A^{B,C}$
is a complex constant with dimensions of length that is completely
determined by short-distance factors:
\begin{eqnarray}
c_A^{B,C}  &=& -i\frac{M_{12} ({\cal C}_A^{B,C}
    - i \lambda_0 {\cal C}_A^{B,12} {\cal C}_{12}^C)}
        {2 \pi {\cal C}_A^{B,12} {\cal C}_{12}^C} .
\label{c:ABC}
\end{eqnarray}
The nonresonant term $c_A^{B,C}$ in Eq.~(\ref{T:ABC-simple})
is a combination of short-distance factors,
so the natural scale for $c_A^{B,C}$ is $1/m$.
The condition $\kappa_X \ll m$ and the condition on $M$ in
Eq.~(\ref{E-small}) imply that the nonresonant term in
Eq.~(\ref{T:ABC-simple}) is small compared to the resonant term
in the threhsold region secified by Eq.~(\ref{E-small}).
We therefore set $c_A^{B,C}=0$.

The invariant mass distribution of the particles in $C$
is obtained by squaring the T-matrix element and integrating
over the momenta of all the particles in the final state.
If $A$ consists of a single heavy particle, the invariant-mass
distribution of the particles in $C$ has the form
\begin{eqnarray}
\frac{d\Gamma}{dM}[A \to B + C] &=&
\big( \Gamma_A^B \Gamma^C \big) \, \frac{M_{12}^2}{2 \pi}
\left| \mathcal{A}_0(M) \right|^2.
\label{dGamABC}
\end{eqnarray}
The short-distance factors $\Gamma_A^B$ and $\Gamma^C$
are the same as in Eqs.~(\ref{GamABX}) and (\ref{GamXCM-fact}).
At 0$^{\rm th}$ order in $g$, $\mathcal{A}_0(E)$ is given in
Eq.~(\ref{A0-ren}). At LO in the complex mass scheme,
$\mathcal{A}_0(E)$ is given in Eq.~(\ref{A0-cms}).
Note that if the invariant mass distribution in Eq.~(\ref{dGamABC})
is divided by the product of the decay rates $\Gamma[A \to B + X]$
and $\Gamma[X \to C]$ in Eqs.~(\ref{GamABX}) and (\ref{GamXCM-fact}),
the short-distance factors cancel.  This combination of observables
has only a long-distance factor given by
$| \mathcal{A}_0(M)|^2/(2 \pi |Z_X|^2)$.

\begin{figure}[t]
\includegraphics[width=12cm]{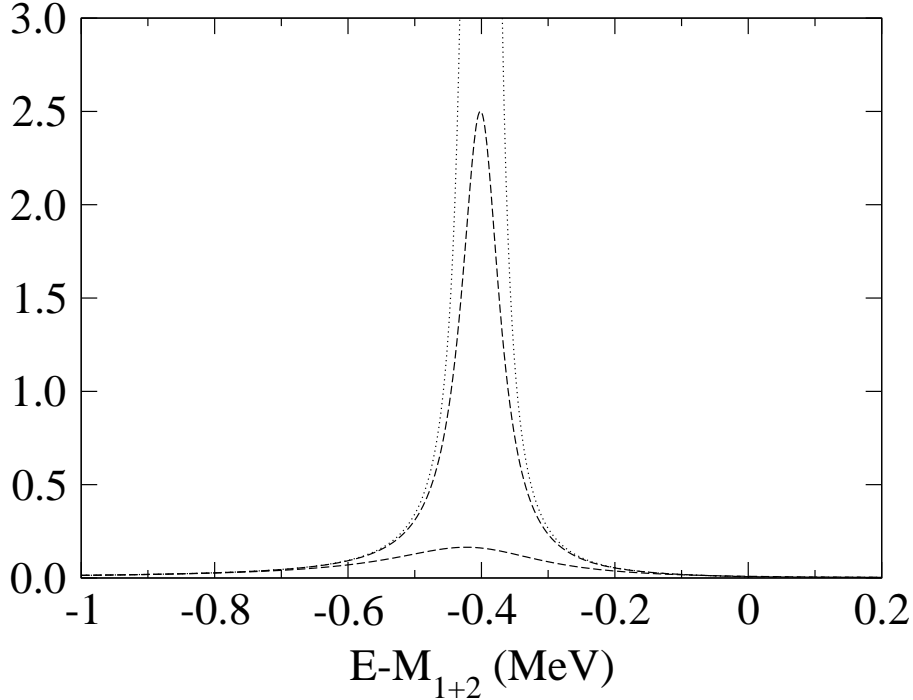}
\caption{
Line shapes $|\mathcal{A}_0(E)|^2$ of the $X$ resonance in a
short-distance decay channel at leading order in $g$.
The dotted line is at order $g^0$ in ordinary perturbation theory
(using $\mathcal{A}_0(E)$ in Eq.~(\ref{A0-ren})).
The dashed lines are at LO in the complex mass scheme
(using $\mathcal{A}_0(E)$ in Eq.~(\ref{A0-cms})).
The line shapes are shown for $\kappa_X = 27.8$ MeV
and for both $\Gamma_2 = 0.07$ MeV (upper solid line)
and $\Gamma_2 = 0.28$ MeV (lower solid line).
The units on the vertical axis are $10^{-5} \ {\rm MeV}^{-4}$.
\label{fig:lineshapeLO}}
\end{figure}

The line shape of $X$ as a function of the invariant mass $E$
of the particles in $C$ is given by the factor $| \mathcal{A}_0(E)|^2$
in Eq.~(\ref{dGamABC}).  In Fig.~\ref{fig:lineshapeLO},
we compare the line shapes at 0$^{\rm th}$ order in $g$
and at LO in the complex mass scheme.
At 0$^{\rm th}$ order in $g$, the line shape diverges at
$M_{1+2} - \kappa_X^2/(2 M_{12})$.
In the complex mass scheme, the line shape is a resonance whose
maximum is near $M_{1+2} - \kappa_X^2/(2 M_{12})$
and whose width is determined by $\Gamma_2$.
If $\Gamma_2 \ll \kappa_X^2/(2 M_{12})$, the line shape
for $|E - M_{1+2} + \kappa_X^2/(2 M_{12})| \ll \kappa_X^2/(2 M_{12})$
is a nonrelativistic Breit-Wigner resonance whose full width
at half-maximum is $\Gamma_2$:
\begin{eqnarray}
\left| \mathcal{A}_0(E) \right|^2 \approx
\frac{4 \pi^2 \kappa_X^2/M_{12}^4}
    {[E - M_{1+2} + \kappa_X^2/(2 M_{12})]^2 + \Gamma_2^2/4}.
\end{eqnarray}
However, a Breit-Wigner resonance has tails that fall
off like $E^{-2}$ and it is integrable.  In contrast, the line shape
$| \mathcal{A}_0(E)|^2$ with $\mathcal{A}_0(E)$ given by
Eq.~(\ref{A0-cms}) has tails that fall
off only like $E^{-1}$ and it is therefore not integrable.
Thus the area under the resonance is not as well-defined as for a
Breit-Wigner resonance.

\section{Second Order in $\bm{g}$}
\label{sec:second}

In this section, we calculate the results in
Section~\ref{sec:zeroth} to next-to-leading order in $g$.
Since the $D_2 - D_1 \phi$ interaction allows transitions to
$D_1 D_1 \phi$ states, we also calculate the rates
for processes whose final state includes $D_1 D_1 \phi$.

\subsection{Amplitude for $\bm{D_1 D_2 \to D_1 D_2}$}
\label{sec:Xamp2}

One of the basic building blocks for the amplitudes for processes
near the $D_1D_2$ threshold is the amplitude
for the propagation of $D_1D_2$ between contact interactions.
The term in this amplitude of order $g^0$, $i L_0(E)$,
is given in Eq.~(\ref{L0-sub}).
The term of order $g^2$, $i L_2(E)$, is the sum of the three
Feynman diagrams in Fig.~\ref{fig:L-NLO}.
These diagrams are calculated in Section~\ref{sec:Propcon2}
of the Appendix.
The final expression for $L_2(E)$ is the sum of the amplitudes
$L_{2a}(E)$ and $L_{2b}(E)$ given by Eqs.~(\ref{L3-integral})
and (\ref{L2-analytic}):
\begin{equation}
L_2(E) = L_2(2M_1 + m) + \frac{M_{12}}{2 \pi} F(\kappa(E)),
\label{L2-bare}
\end{equation}
where $\kappa(E)$ is the energy variable defined in Eq.~(\ref{kappa-E}),
the function $F(\kappa)$ is
\begin{eqnarray}
F(\kappa) &=&
\frac{g^2 M_{12} m_1}{\pi^2}
\Bigg[ 2 \int_0^1 \frac{dz}{z^2}
\left( \frac{t^3(z)}{1-t(z)} \right)^{1/2}
\ln \frac{z \kappa^2 + (1-z) (m_1/M_{12}) (\kappa^2 - \kappa_1^2)}
    {z \kappa_1^2}
\nonumber
\\
&& \hspace{2cm}
- \left( \frac{m_1}{M_{12}} \right)^{1/2} \,
\left( 2 \ln \frac{\kappa+\kappa_1}{2 \kappa_1}
+ \frac{(\kappa - \kappa_1)^2}{2 \kappa_1 \kappa}
\ln \frac{\kappa + \kappa_1}{\kappa - \kappa_1} \right) \Bigg],
\label{F-kappa}
\end{eqnarray}
and $\kappa_1$ is the value of $\kappa(E)$ at the $D_1 D_1 \phi$
threshold $2 M_1 + m$:
\begin{eqnarray}
\kappa_1 &=& [ 2M_{12}(M_{2-1}-m)]^{1/2}.
\label{kappa1}
\end{eqnarray}
The function $t(z)$ in the integrand in Eq.~(\ref{F-kappa})
is a rational function of $z$ that increases from 0 to 1
as $z$ increases from 0 to 1:
\begin{equation}
t(z) = \frac{z^2}{(2-z)^2 - 4 (m_1/m)^2(1-z)^2}.
\label{t-z}
\end{equation}
The term $L_2(2M_1 + m)$ in Eq.~(\ref{L2-bare}) is real valued
and includes logarithmic ultraviolet divergences.
The function $F(\kappa)$ vanishes at the $D_1 D_1 \phi$ threshold
$\kappa = \kappa_1$ by definition.
The first few terms in the Laurent expansion of $F(\kappa)$
around the $D_1 D_2$ threshold at $\kappa = 0$ can be obtained
from Eqs.~(\ref{L2a-Laurent}) and (\ref{L2b-Laurent}).
The leading term is
\begin{equation}
F(\kappa) \longrightarrow
- i \frac{g^2 M_{12}^2 (m_1/M_{12})^{3/2} \kappa_1}{2 \pi \kappa}.
\label{F-Laurent}
\end{equation}

\begin{figure}[t]
\includegraphics[width=12cm]{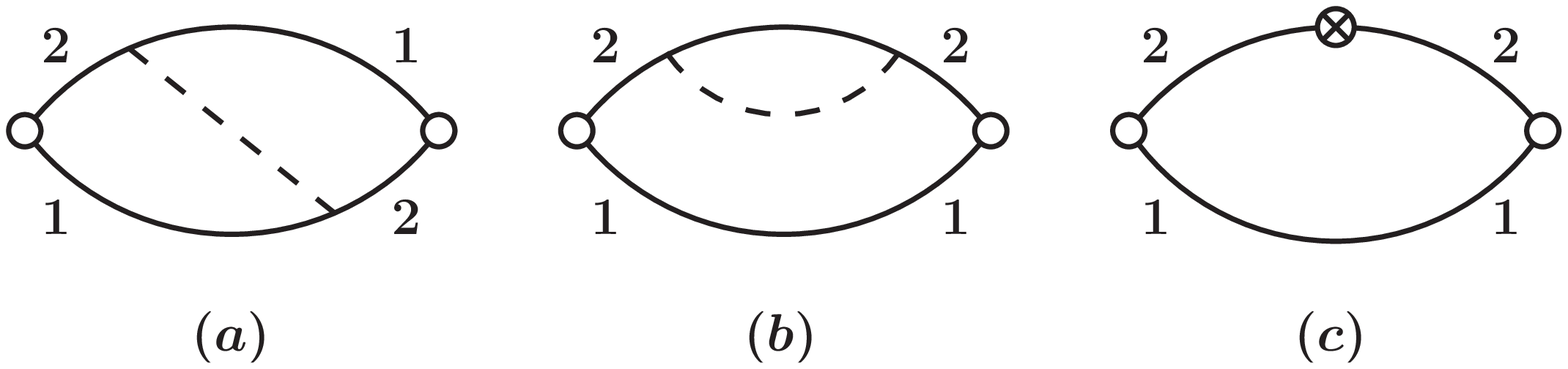}
\caption{ Diagrams of order $g^2$ for the amplitude for the
propagation of $D_1D_2$ between contact interactions.
The open dots indicate that the vertex factors $- i \lambda_0$
are omitted.
\label{fig:L-NLO}}
\end{figure}

At 0$^{\rm th}$ order in $g$, the ultraviolet divergence in $L_0(E)$
was eliminated by summing a geometric series in $L_0$ and then
renormalizing the coupling constant $\lambda_0$.
The ultraviolet divergence in $L_2(E)$ can be eliminated in a
similar way.
The geometric series of diagrams that must be summed are those
that can be obtained by replacing each one-loop subdiagram in
Fig.~\ref{fig:ALO} by the sum of the one-loop subdiagram
and the three diagrams for $L_2(E)$ in Fig.~\ref{fig:L-NLO}.
The sum of that geometric series gives an amplitude
$i\mathcal{A}_2(E)$ that can be obtained by making the substitution
$L_0 \to L_0+ L_2$ in Eq.~(\ref{A-bare}):
\begin{equation}
i\mathcal{A}_2(E) =
\frac{-i}{1/\lambda_0 - [L_0(E) + L_2(E)]}.
\label{A2-bare}
\end{equation}
Our renormalization prescription for $\lambda_0$
in Eq.~(\ref{ReAinv:zero}) requires ${\rm Re} \, \mathcal{A}_2^{-1}(E)$
to vanish at $M_{1+2} - \kappa_X^2/(2 M_{12})$.
Using this prescription,
the amplitude in Eq.~(\ref{A2-bare}) can be expressed as
\begin{eqnarray}
\mathcal{A}_2(E) &=&
\frac{2 \pi/M_{12}}
    { \kappa(E) + F(\kappa(E))
    - {\rm Re} \, [ \kappa_X +  F(\kappa_X - i \epsilon) ]} .
\label{A2-NLO}
\end{eqnarray}

\begin{figure}[t]
\includegraphics[width=12cm]{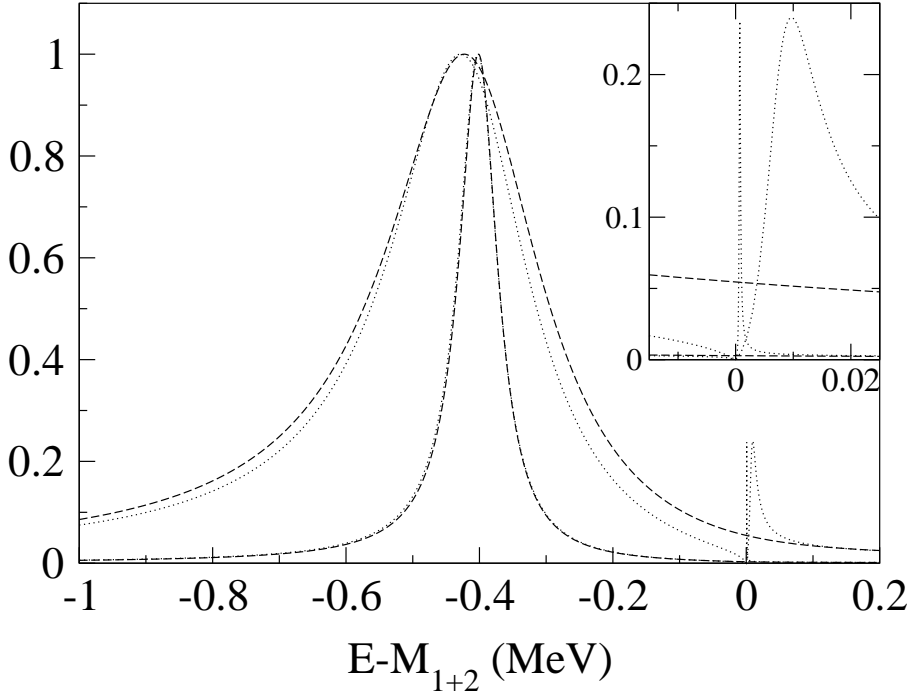}
\caption{
Line shapes $|\mathcal{A}(E)|^2$ of the $X$ resonance in a short-distance
decay channel.
The dotted lines are at order $g^2$ in ordinary perturbation theory
(using $\mathcal{A}_2(E)$ in Eq.~(\ref{A2-NLO})).
The dashed lines are at LO in the complex mass scheme
(using $\mathcal{A}_0(E)$ in Eq.~(\ref{A0-cms})).
The line shapes are shown for $\kappa_X = 27.8$ MeV
and for both $\Gamma_2 = 0.07$ MeV (the two narrower resonances
that are almost indistinguishable)
and for $\Gamma_2 = 0.28$ MeV (the two wider resonances).
The line shapes are normalized so their maximum values are 1. The inset
shows their behavior near the $D_1 D_2$ threshold.
\label{fig:lineshapeNLOcmsLO}}
\end{figure}

The quantity $|\mathcal{A}_2(E)|^2$ with $\mathcal{A}_2(E)$
given in Eq.~(\ref{A2-NLO}) is the line shape of the $X$ resonance
in a short-distance decay channel to order $g^2$
in ordinary perturbation theory.
In Fig.~\ref{fig:lineshapeNLOcmsLO}, we compare this line shape
with that at LO in the complex mass scheme, which is
$|\mathcal{A}_0(E)|^2$ with $\mathcal{A}_0(E)$ given by
Eq.~(\ref{A0-cms}).   They are shown for two values of the
coupling constant $g$ that correspond to the width of $D_2$ being
$\Gamma_2 = 0.07$ MeV and $0.28$ MeV.
The two line shapes have the same qualitative
behavior below the resonance and near the peak of the resonance.
They have qualitatively
different behavior near the $D_1 D_2$ threshold $E= M_{1+2}$.
Unlike $|\mathcal{A}_0(E)|^2$, the function $|\mathcal{A}_2(E)|^2$
vanishes at the $D_1 D_2$ threshold
and has a sharp peak just above the threshold.  The inset in
Fig.~\ref{fig:lineshapeNLOcmsLO} shows the behavior near the
$D_1 D_2$ threshold, which is clearly unphysical.

The reason $|\mathcal{A}_2(E)|^2$ vanishes at the $D_1 D_2$ threshold
can be seen from the Laurent expansion of $F(\kappa)$
around $\kappa = 0$.
The leading term, which is given in Eq.~(\ref{F-Laurent}),
diverges like $1/\kappa$ as $E \to M_{1+2}$.
No matter how small the coupling constant $g$, the term proportional to
$g^2/\kappa$ cannot be
treated as a perturbation in the region near $\kappa = 0$.
The resolution of the problem can be found by noting that
the diverging term in Eq.~(\ref{F-Laurent}) can be expressed as
\begin{eqnarray}
 F(\kappa)  &\longrightarrow&
- i \frac{M_{12} \Gamma_2}{2 \kappa} ,
\end{eqnarray}
where $\Gamma_2$ is the width of the $D_2$ from its decay
into $D_1 \phi$, which is given in Eq.~(\ref{Gam2-NR}).
Thus the divergence at $\kappa = 0$ implies that near the $D_1 D_2$
threshold, the imaginary part of the $D_2$ self-energy
must be resummed to all orders.  That resummation shifts
the branch point in the leading order amplitude
from the real threshold $E = M_{1+2}$ to the complex
threshold $E = M_{1+2}- i \Gamma_2/2$.
Making this change and then imposing our renormalization scheme,
the amplitude in Eq.~(\ref{A2-NLO}) becomes
\begin{eqnarray}
\mathcal{A}_2(E) \approx
\frac{2 \pi/M_{12}}
    { \gamma(E)
    + F(\kappa(E))  + i M_{12} \Gamma_2/(2 \kappa)
    - {\rm Re} [ \gamma_X + F(\kappa_X - i \epsilon)] } ,
\label{A2-resum}
\end{eqnarray}
where $\gamma(E)$ and $\gamma_X$ are defined in Eqs.~(\ref{gamma-E})
and (\ref{gammaX}).
This amplitude is a smooth function of $E$ at $\kappa = 0$.

A convenient way to implement the resummation in
Eq.~(\ref{A2-resum}) that can be extended staightforwardly
to higher orders in $g$ is to use the complex mass scheme.
This automatically gives
an amplitude $\mathcal{A}_2(E)$ that has smooth behavior
at the $D_1D_2$ threshold and is correct through relative order $g^2$.
In the complex mass scheme, the contribution to $L_2(E)$
from the diagrams in Figs.~\ref{fig:L-NLO}(a) and \ref{fig:L-NLO}(b)
are obtained by replacing the energy variable $\kappa(E)$ by
$\gamma(E)$ defined in Eq.~(\ref{gamma-E}) and by replacing
$\kappa_1$ by $\gamma_1$, which is the value of $\gamma(E)$ at the
$D_1 D_1 \phi$ threshold $2 M_1+m$:
\begin{eqnarray}
\gamma_1 &=& \left[ 2 M_{12} (M_{2-1} - m- i \Gamma_2/2) \right]^{1/2} .
\label{gamma1}
\end{eqnarray}
In the complex mass scheme, there are additional contributions
to the function $L_2(E)$ coming from the one-loop diagram
in Fig.~\ref{fig:L-NLO}(c) with the counterterm vertex in
Eq.~(\ref{DLint}).  The additional contributions
are given in Eq.~(\ref{L2c}).
Adding those terms is equivalent to making the substitution
\begin{equation}
F(\kappa)  \longrightarrow
F(\kappa) + \frac{i M_{12} \Gamma_2}{2 \kappa}
+ (Z_2^{-1} - 1) \left( \frac{2\pi}{M_{12}} L_0(M_{1+2}) + \kappa \right) .
\label{Fsub}
\end{equation}
Using the expressions to order $g^2$ for $\Gamma_2$ in
Eq.~(\ref{Gam2-NR}) and $Z_2^{-1}$ in Eq.~(\ref{Z2-NR}),
we can see that the additional terms in Eq.~(\ref{Fsub})
cancel the $\kappa^{-1}$ and $\kappa$ terms coming from
the Laurent expansions of $L_{2b}(E)$ in Eq.~(\ref{L2b-Laurent}).
Our renormalization prescription for $\lambda_0$
in Eq.~(\ref{ReAinv:zero}) requires ${\rm Re} \, \mathcal{A}_2^{-1}(E)$
to have a zero at $M_{1+2} - \kappa_X^2/(2 M_{12})$.
Using this prescription,
the amplitude can be expressed as
\begin{eqnarray}
\mathcal{A}_2(E) &=&
\frac{2 \pi/M_{12}}
    { \gamma(E) + F_{\rm cms}(\gamma(E))
    - {\rm Re} \, [ \gamma_X +  F_{\rm cms}(\gamma_X) ]},
\label{A2-ren:resum}
\end{eqnarray}
where $\gamma(E)$ and $\gamma_X$ are given by Eqs.~(\ref{gamma-E})
and (\ref{gammaX}) and the function $F_{\rm cms}(\gamma)$ is
\begin{eqnarray}
F_{\rm cms}(\gamma) &=&
\frac{g^2 M_{12} m_1}{\pi^2}
\Bigg[ 2 \int_0^1 \frac{dz}{z^2}
\left( \frac{t^3(z)}{1-t(z)} \right)^{1/2}
\left( \ln \frac{(1-z) (m_1/M_{12}) (\gamma_1^2 - \gamma^2) - z \gamma^2}
       {(1-z) (m_1/M_{12}) \gamma_1^2}
- i \pi \right)
\nonumber
\\
&& \hspace{2cm}
- \left( \frac{m_1}{M_{12}} \right)^{1/2} \,
\left( 2 \ln \frac{\gamma_1+\gamma}{\gamma_1}
+ \frac{(\gamma_1 - \gamma)^2}{2 \gamma_1 \gamma}
\ln \frac{\gamma_1 + \gamma}{\gamma_1 - \gamma} - 1 - i \pi \right) \Bigg].
\label{Fcms}
\end{eqnarray}
The function $F_{\rm cms}(\gamma)$ can be obtained from $F(\kappa)$
in Eq.~(\ref{F-kappa}) by subtracting several terms in the Laurent
expansion in $\kappa$ and then replacing $\kappa(E)$ and $\kappa_1$
by $\gamma(E)$ and $\gamma_1$.  The terms that are subtracted are
the imaginary term of order $\kappa^{-1}$,
the real terms of order $\kappa^0$,
and the imaginary term of order $\kappa$ in the second of the two terms
on the right side of Eq.~(\ref{F-kappa}).
The real terms of order $\kappa^0$ are subtracted because
they would cancel in the denominator in Eq.~(\ref{A2-ren:resum}) anyway.
The leading terms in the Laurent expansion of $F_{\rm cms}(\gamma)$
around the complex $D_1 D_2$ threshold at $\gamma = 0$ are
\begin{eqnarray}
F_{\rm cms}(\gamma) & \longrightarrow&
\frac{ig^2 M_{12} m_1}{\pi}
\Bigg[ - \frac{m}{m_1} \arcsin \frac{m_1}{m}
+ \left( \frac{m_1}{M_{12}} \right)^{1/2}
+ 2 \left( \frac{M_{12}}{m_1} \right)^{1/2} \frac{\gamma}{\gamma_1} \Bigg].
\end{eqnarray}
For real values of $E$, there is an ambiguity in the function 
$F_{\rm cms}(\gamma(E))$ in Eq.~(\ref{A2-ren:resum}) for $E < 2M_1+m$
because of the branch point at $\gamma = \gamma_1$ in Eq.~(\ref{Fcms}).
This ambiguity can be resolved by the substitution
$\gamma \to \gamma - i \epsilon$.

\begin{figure}[t]
\includegraphics[width=12cm]{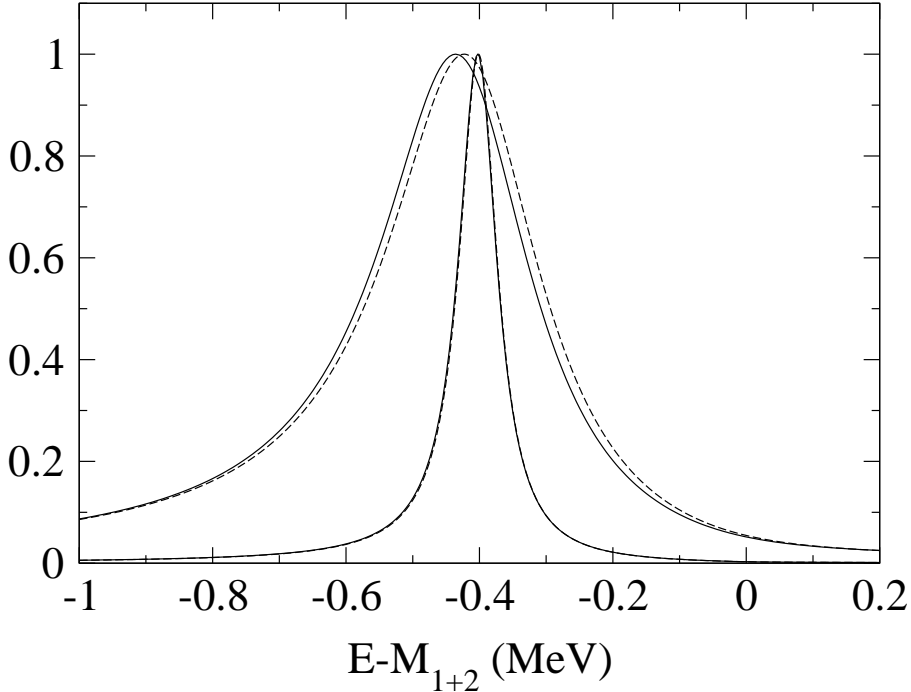}
\caption{
Line shapes $|\mathcal{A}(E)|^2$ of the $X$ resonance
in a short-distance decay channel using the complex mass scheme.
The dashed lines are at LO in the complex mass scheme
(using $\mathcal{A}_0(E)$ in Eq.~(\ref{A0-cms})).
The solid lines are at NLO in the complex mass scheme
(using $\mathcal{A}_2(E)$ in Eq.~(\ref{A2-ren:resum})).
The line shapes are shown for $\kappa_X = 27.8$ MeV
and for both $\Gamma_2 = 0.07$ MeV (the two narrower resonances
that are almost indistiguishable)
and for $\Gamma_2 = 0.28$ MeV (the two wider resonances).
The line shapes are normalized so their maximum values are 1.
\label{fig:lineshapeNLOcmsNLO}}
\end{figure}

In Fig.~\ref{fig:lineshapeNLOcmsNLO}, we compare the line shapes
of $X$ in a short-distance decay channel
at NLO in the complex mass scheme, which is $|\mathcal{A}_2(E)|^2$
with $\mathcal{A}_2(E)$ given by Eq.~(\ref{A2-ren:resum}), and
that at LO in the complex mass scheme, which is
$|\mathcal{A}_0(E)|^2$ with $\mathcal{A}_0(E)$ given by
Eq.~(\ref{A0-cms}).  They are shown for two values of the
coupling constant $g$ that correspond to the width of $D_2$ being
$\Gamma_2 = 0.07$ MeV and $0.28$ MeV.
The curves are normalized so their maximum values are 1.
The line shapes at LO and NLO have the same qualitative behavior.
As $g$ decreases, the quantitative differences between the line shapes
at LO and NLO decrease.  Thus the complex mass scheme seems to
provide a good solution to the infrared problem near the
$D_1 D_2$ threshold that arises in ordinary perturbation theory.
We will therefore use the complex mass scheme in most of the
calculations in the remainder of this Section.

\subsection{Propagator for $\bm{X}$}
\label{sec:Xprop2}

If the local composite operator $\lambda_0 D_1 D_2(x)$ is used as the
interpolating field for $X$,
the propagator for $X$ is defined in Eq.~(\ref{X-prop}).
The expression for the propagator at 0$^{\rm th}$ in $g$
is given in Eq.~(\ref{DeltaX-0}).
A renormalizable expression for the $X$ propagator that is
accurate to order $g^2$ can be obtained simply by making the
substitution $L_0 \to L_0 + L_2$ in Eq.~(\ref{DeltaX-0}):
\begin{equation}
\Delta_X(E,0) =
\frac{i \lambda_0^2 [L_0(E) + L_2(E)]}
    {1 - \lambda_0 [L_0(E) + L_2(E)]} .
\label{DeltaX-2}
\end{equation}
Comparing with the expression for the amplitude $\mathcal{A}_2(E)$ in
Eq.~(\ref{A2-bare}), we see that the propagator
in Eq.~(\ref{DeltaX-2}) can be expressed as
\begin{equation}
\Delta_X(E,0) = -i[\lambda_0 + \mathcal{A}_2(E)].
\label{DeltaX-2ren}
\end{equation}

The expression for $\mathcal{A}_2(E)$ at NLO in the complex mass scheme
is given in Eq.~(\ref{A2-ren:resum}).  This amplitude has a pole
at a complex energy $E_{\rm pole}$ that determines the binding energy
and the width of $X$.
The energy $E_{\rm pole}$ is the solution to
\begin{equation}
\gamma(E_{\rm pole})
+ F_{\rm cms}(\gamma(E_{\rm pole}))
    - {\rm Re} \, [ \gamma_X +  F_{\rm cms}(\gamma_X) ] = 0,
\label{gamma-pole:NLO}
\end{equation}
where $\gamma(E)$ and $\gamma_X$ are defined in
Eqs.~(\ref{gamma-E}) and (\ref{gammaX}).
The wavefunction normalization constant for $X$ is
\begin{equation}
Z_X = \frac{2 \pi}{M_{12}^2}
\frac{ {\rm Re} \, \gamma(E_{\rm pole})}
    {1+ F_{\rm cms}'(\gamma(E_{\rm pole}))} .
\label{ZX-NLO}
\end{equation}

If we treat $\Gamma_2$ as order $g^2$,
the solution to Eq.~(\ref{gamma-pole:NLO})
for the complex variable $E_{\rm pole}$ through order $g^2$ is
\begin{equation}
E_{\rm pole} \approx M_{1+2} - \kappa_X^2 /(2 M_{12})
- i \left[  \Gamma_2/2
    - \kappa_X \, {\rm Im} \, F_{\rm cms}(\kappa_X)/M_{12} \right].
\end{equation}
Using the expression for $F_{\rm cms}(\gamma)$ in Eq.~(\ref{Fcms}),
we find that the binding energy and width of the $X$ to order $g^2$ are
\begin{subequations}
\begin{eqnarray}
E_X &\approx& \kappa_X^2/(2 M_{12}),
\\
\Gamma_X &\approx& \Gamma_2
+ \frac{2 g^2 m_1 \kappa_X}{\pi}
\left[ 2 \int_0^{z_0} \frac{dz}{z^2} \left( \frac{t^3(z)}{1-t(z)} \right)^{1/2}
    - \left( \frac{m_1}{M_{12}} \right)^{1/2} \right],
\label{GamX:NLO}
\end{eqnarray}
\label{EGamX:NLO}
\end{subequations}
where the upper endpoint $z_0$ of the integral is
\begin{equation}
z_0 =
\frac{(m_1/M_{12})(\kappa_1^2 - \kappa_X^2)}
    {(m_1/M_{12})(\kappa_1^2 - \kappa_X^2) + \kappa_X^2}.
\end{equation}
For $\kappa_X = 27.8$ MeV, the second term on the right side of
Eq.~(\ref{GamX:NLO}) is $0.011 \, \Gamma_2$.
Thus the width of the constituent $D_2$ accounts for most of the
width of $X$ at order $g^2$.

We can obtain an analytic expression for the width $\Gamma_X$
if $\kappa_X \ll (m_1/M_{12})^{1/2} \kappa_1$.
In this limit, the upper limit $z_0$ on
the integral approaches 1 and the width reduces to
\begin{equation}
\Gamma_X \approx \Gamma_2
+ \frac{2 g^2 m_1 \kappa_X}{\pi}
\left[ \frac{m}{m_1} \arcsin \frac{m_1}{m}
    - \left( \frac{m_1}{M_{12}} \right)^{1/2} \right].
\label{gamma-X:NLOlim}
\end{equation}
As $m/M_1$ varies from 0 to $\infty$, the factor in square brackets
ranges from $\pi/2$ to $1-\sqrt{M_{1+2}/M_2}$.
Note that the expression for $\Gamma_X$ in Eq.~(\ref{gamma-X:NLOlim})
can be smaller than $\Gamma_2$ for some values of the parameters.

\subsection{Short-distance Production of $\bm{X}$}
\label{sec:Xprod2}

The operator product expansion of the T-matrix element
for the short-distance production process $A \to B + X$
is given in Eq.~(\ref{TAXpB}).
At NLO in the complex mass scheme,
the operator matrix element is given by Eq.~(\ref{XDD0}),
where $Z_X$ is now the wavefunction normalization constant for
$X$ in Eq.~(\ref{ZX-NLO}).  The factored expression for the T-matrix element
is given by Eq.~(\ref{TAXB-fact}). If $A$ consists of a single particle,
its decay rate into $B+X$ can be expressed in the factored form in
Eq.~(\ref{GamABX}), where $Z_X$ is given in Eq.~(\ref{ZX-NLO}).

\subsection{Short-distance decay of $\bm{X}$}
\label{sec:Xdecay2}

The operator product expansion for the T-matrix element
for the short-distance decay process $X \to C$
is given in Eq.~(\ref{TXCM}).
At NLO in the complex mass scheme,
the operator matrix element is
given in Eq.~(\ref{0DDX}), where $Z_X$ is the wavefunction
normalization constant for $X$ in Eq.~(\ref{ZX-NLO}).
The factored expression for the T-matrix element
is given by Eq.~(\ref{TXCM-fact}).  The decay rate of $X$ into
the particles represented by $C$ can be expressed in the
factored form in Eq.~(\ref{GamXCM-fact}),
where $Z_X$ is given in Eq.~(\ref{ZX-NLO}).

\subsection{Line shape of $\bm{X}$ in a Short-distance Decay Channel}
\label{sec:Xlineshape2}

The T-matrix element for the short-distance process $A \to B + C$,
where $C$ is a set of particles with invariant mass $M$
close to the $D_1 D_2$ threshold, can be expressed as the double
operator product expansion in Eq.~(\ref{T:ABCM}).
The T-matrix element at order $g^0$
is given in Eq.~(\ref{T:ABC-bare}).  An
expression for this T-matrix element that is accurate through
order $g^2$ can be obtained by substituting
$L_0 \to L_0 + L_2$:
\begin{eqnarray}
{\cal T}[A \to B + C]  &=& {\cal C}_A^{B,C} +
{\cal C}_A^{B,12} {\cal C}_{12}^C
\frac{i \lambda_0^2 [L_0(M) + L_2(M)]}
    {1 - \lambda_0 [L_0(M) + L_2(M)]} .
\label{T:ABC-bare-NLO}
\end{eqnarray}
The expression for the T-matrix element in which short-distance
and long-distance factors are separated is given by Eq.~(\ref{T:ABC-simple})
with $\mathcal{A}_0(M)$ replaced by $\mathcal{A}_2(M)$.
The invariant mass distribution for the short-distance decay channel is
\begin{eqnarray}
\frac{d\Gamma}{dM}[A \to B + C] &=&
\big( \Gamma_A^B \Gamma^C \big) \, \frac{M_{12}^2}{2 \pi}
\left| \mathcal{A}_2(M) \right|^2,
\label{dGamABC:NLO}
\end{eqnarray}
where $\Gamma_A^B$ and $\Gamma^C$ are the short-distance factors
defined in Eqs.~(\ref{GamADDB}) and (\ref{Gam-C}).
The expression for $\mathcal{A}_2(M)$ at NLO in the complex mass scheme
is given in Eq.~(\ref{A2-ren:resum}).
The line shapes at LO and NLO in the complex mass scheme are
compared in Fig.~\ref{fig:lineshapeNLOcmsNLO}.

If the invariant mass distribution in Eq.~(\ref{dGamABC:NLO})
is divided by the product of the decay rates $\Gamma[A \to B + X]$
and $\Gamma[X \to C]$ in Eqs.~(\ref{GamABX}) and (\ref{GamXCM-fact}),
the short-distance factors cancel, leaving only the long-distance
factors $|\mathcal{A}_2(M)|^2/(2 \pi |Z_X|^2)$,
where $\mathcal{A}_2(E)$ and $Z_X$ are given in Eqs.~(\ref{A2-ren:resum})
and (\ref{ZX-NLO}).  The corresponding quantity at LO in the
complex mass scheme is $|\mathcal{A}_0(M)|^2/(2 \pi |Z_X|^2)$,
where $\mathcal{A}_0(E)$ and $Z_X$ are given in Eqs.~(\ref{A0-cms})
and (\ref{ZX:cms}).
The dependence of this long-distance factor on the invariant mass $M$
is illustrated in Fig.~\ref{fig:lineshapeNLOcmsNLO},
where the curves are all normalized to maximum value 1.
The peak value of $|\mathcal{A}(M)|^2/(2 \pi |Z_X|^2)$
is also completely determined by long distances.
For $\kappa_X = 27.8$ MeV and $\Gamma_2 = 0.07$ MeV, the peak value
decreases from 130.1 MeV$^{-2}$ at LO to 119.2 MeV$^{-2}$ at NLO.
For $\kappa_X = 27.8$ MeV and $\Gamma_2 = 0.28$ MeV, the peak value
decreases from 8.34 MeV$^{-2}$ at LO to 6.58 MeV$^{-2}$ at NLO.
The decrease in the difference between the peak values at NLO and LO
from about 27\% at $\Gamma_2 = 0.28$ MeV to about 9\% at $\Gamma_2 = 0.07$ MeV
supports the assumption that the
perturbative expansion is well-behaved in the complex mass scheme.

\subsection{Short-distance production of $\bm{D_1 D_1 \phi}$}
\label{sec:DDphi1}

The fundamental theory may have short-distance production processes
$A \to B + D_1 D_1 \phi$, where $A$ and $B$ both represent
one or more particles whose momenta in the $D_1 D_1 \phi$ rest
frame are of order $m$ or larger.
The $D_1 D_1 \phi$ invariant mass $M$ is assumed to be close to $M_{1+2}$,
as specified by the condition in Eq.~(\ref{E-small}).
The operator product expansion of the T-matrix element
for this process is given in Eq.~(\ref{TADDphiB0}).

\begin{figure}[t]
\includegraphics[width=12cm]{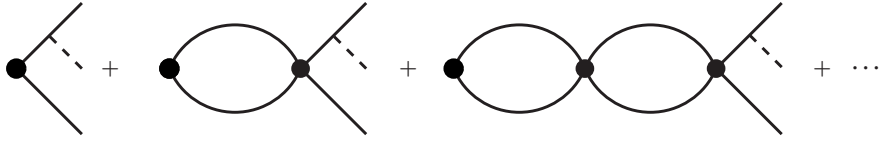}
\caption{
Diagrams at order $g$ for the vacuum--to--$D_1D_2\phi$
matrix element of the operator $\lambda_0 D_1^\dagger D_2^\dagger$.
\label{fig:DDphi}}
\end{figure}

We first calculate the rate for $A \to B + D_1 D_1 \phi$
to order $g^2$ using ordinary perturbation theory.
At order $g$,
the operator matrix element in Eq.~(\ref{TADDphiB0}) is given by
the series of Feynman diagrams in Fig.~\ref{fig:DDphi} summed over the two
permutations of the external $D_1$ lines.  The sum of the
geometric series of diagrams is $\mathcal{A}_0(M)/(-\lambda_0)$
multiplied by the first diagram in the series.  According to the
Optical Theorem, the square of the first diagram integrated over the
3-body phase space for $D_1 D_1 \phi$ is given by
the sum of the 3-body cuts through the Feynman diagrams
for $L_2(E)$ in Fig.~\ref{fig:L-NLO}.  If $M < M_{1+2}$,
the sum of the 3-body cuts gives the imaginary part of $L_2(E)$.
Thus the square of the T-matrix element in Eq.~(\ref{TADDphiB0})
integrated over the 3-body phase space of $D_1 D_1 \phi$ is
\begin{eqnarray}
&& \frac{1}{2} \int \frac{d^3p_1}{(2 \pi)^3 2E_1}
    \frac{d^3p_2}{(2 \pi)^3 2 E_2} \frac{d^3k}{(2 \pi)^3 2 \omega}
(2 \pi)^4 \delta(P - p_1 -p_2 - k)
\left| {\cal T}[A \to B + D_1 D_1 \phi] \right|^2
\nonumber
\\
&& \hspace{7cm}
= |{\cal C}_A^{B,12} |^2 |\mathcal{A}_0(M)|^2
(-2) \, {\rm Im} \, L_2(M) ,
\label{TDDphi:g2}
\end{eqnarray}
where $p_1$, $p_2$, and $k$ are 4-momenta for $D_1$, $D_1$, and $\phi$,
respectively, and $P^2= M^2$.
If $A$ consists of a single particle,
the invariant mass distribution for $A \to B + D_1D_1\phi$
at order $g^2$ can be expressed as
\begin{eqnarray}
\frac{d\Gamma}{dM}[A \to B + D_1 D_1 \phi]
&=& \Gamma_A^B \,
\frac{M_{12}}{2\pi}
\left| \mathcal{A}_0 (M)\right|^2
(-2) \, {\rm Im} \, L_2(M) ,
\label{DDphiIMD:g2}
\end{eqnarray}
where $\Gamma_A^B$ is the short-distance factor defined in
Eq.~(\ref{GamADDB}) and
we have replaced a multiplicative factor of $M$ by $M_{1+2}$.

If there is a short-distance process $A \to B + D_1 D_1 \phi$,
then the process $A \to B + D_1 D_2$ is also allowed.
The invariant mass distribution for $A \to B + D_1 D_2$
at order $g^0$ is given by the expression on the right side of
Eq.~(\ref{DDphiIMD:g2}) with ${\rm Im} \, L_2(M)$ replaced by
${\rm Im} \, L_0(M)$.
The threshold for this process is $M > M_{1+2}$.
Since $D_2$ ultimately decays into
$D_1 \phi$, a decay into $D_1 D_2$ can also be regarded
as a contribution to the inclusive decay into $D_1D_1\phi$.

The result in Eq.~(\ref{DDphiIMD:g2}) is an order $g^2$ correction
to the inclusive invariant mass distribution for $D_1 D_1 \phi$.
It is nonzero for all $M > 2M_1 + m$.
For $M> M_{1+2}$, there are additional corrections of order $g^2$
that can be obtained by replacing any of the one-loop subdiagrams
in Fig.~\ref{fig:DDphi} by one of the two-loop diagrams for $L_2(E)$
in Fig.~\ref{fig:L-NLO}.  Nonperturbative renormalization
of the coupling constant $\lambda_0$ then requires that a
geometric series of these corrections be summed to all orders.
The net effect is to replace $\mathcal{A}_0(M)$ in
Eq.~(\ref{DDphiIMD:g2}) by $\mathcal{A}_2(M)$.
Thus the complete result for the inclusive invariant mass distribution
for $D_1 D_1 \phi$ through order $g^2$ is
\begin{eqnarray}
\frac{d\Gamma}{dM}[A \to B + D_1 D_1 \phi]
&=& \Gamma_A^B \,
\frac{M_{12}}{2\pi}
\left| \mathcal{A}_2 (M)\right|^2
(-2) \, {\rm Im} \, [L_0(M) + L_2(M)] ,
\label{dGamDDphi:g2}
\end{eqnarray}
where the functions $\mathcal{A}_2 (M)$, $L_0(M)$, and $L_2(M)$
are given in Eqs.~(\ref{A2-NLO}), (\ref{L0-sub}), and (\ref{L2-bare}).
Note that taking the imaginary parts of $L_0(M)$ and $L_2(M)$
eliminates the additive ultraviolet-divergent terms.
A more compact expression for the distribution in
Eq.~(\ref{dGamDDphi:g2}) is
\begin{eqnarray}
\frac{d\Gamma}{dM}[A \to B + D_1 D_1 \phi]
&=&
\Gamma_A^B \, \frac{M_{12}}{\pi} \, {\rm Im} \, \mathcal{A}_2(M).
\label{DDphiIMD:NLOcms}
\end{eqnarray}

The inclusive invariant mass distribution
in Eq.~(\ref{dGamDDphi:g2}) when calculated using ordinary perturbation
theory has unphysical behavior at the
threshold $M = M_{1+2}$.  The factor $| \mathcal{A}_2 (M)|^2$
has a zero at $M = M_{1+2}$ and a sharp peak just above that
threshold as illustrated in Fig.~\ref{fig:lineshapeNLOcmsLO}.
The term $-2 \, {\rm Im} L_2(M)$ also has unphysical behavior
at this threshold.
The imaginary part of $L_2(E)$ in the region
$2 M_1 + m < E < M_{1+2}$ is given by the sum of
Eqs.~(\ref{ImL2a}) and (\ref{ImL2b}).
The resulting expression for $-2 \, {\rm Im} \, L_2(E)$ is
\begin{eqnarray}
-2 \, {\rm Im} \, L_2(E) = \frac{g^2 M_{12}^2m_1}{\pi^2}
\left[ 2 \int_0^{z_0} \frac{dz}{z^2}
\left( \frac{t^3(z)}{1-t(z)} \right)^{1/2}
+ \left( \frac{m_1}{M_{12}} \right)^{1/2} \,
\frac{(\kappa_1 - \kappa(E))^2}{2 \kappa_1 \kappa(E)} \right],
\label{ImL2ab}
\end{eqnarray}
where the upper limit $z_0$ of the integral is
\begin{equation}
z_0 =
\frac{(m_1/M_{12})(\kappa_1^2 - \kappa(E)^2)}
    {(m_1/M_{12})(\kappa_1^2 - \kappa(E)^2) + \kappa(E)^2}.
\label{z0-M}
\end{equation}
The function in Eq.~(\ref{ImL2ab})
diverges like $1/\kappa(E)$ as $E \to M_{1+2}$.
As a consequence, near the $D_1 D_2$ threshold, corrections of
higher order in $g^2$ that come from the $D_2$ self energy
are not suppressed.
This is the same problem that prompted us to introduce the
complex mass scheme in Section~\ref{sec:Xamp2}.

We next consider the inclusive rate for $A \to B + D_1 D_1 \phi$
in the complex mass scheme.
At LO in the complex mass scheme,
the invariant mass distribution for $D_1 D_1 \phi$ can be obtained from
Eq.~(\ref{DDphiIMD:NLOcms})
by replacing $\mathcal{A}_2(M)$ by $\mathcal{A}_0(M)$ in
Eq.~(\ref{A0-cms}).  This can be written more explicitly as
\begin{eqnarray}
\frac{d\Gamma}{dM}[A \to B + D_1 D_1 \phi]
&=&
\Gamma_A^B \,
\frac{-2 \, {\rm Im} \, \gamma(M)}{|\gamma(M) - {\rm Re} \, \gamma_X|^2},
\label{DDphiIMD:LOcms}
\end{eqnarray}
where $\gamma(E)$ and $\gamma_X$ are defined in
Eqs.~(\ref{gamma-E}) and (\ref{gammaX}).
This expression has a nonzero imaginary part at all energies $E$,
even below the $D_1 D_1 \phi$ theshold.
For $E < 2 M_1 + m$, the imaginary part is small and it is cancelled
by higher orders in the complex mass scheme.
At NLO in the complex mass scheme,
the invariant mass distribution for $D_1 D_1 \phi$
is obtained from Eq.~(\ref{DDphiIMD:NLOcms})
by using the expression for $\mathcal{A}_2(M)$ in Eq.~(\ref{A2-ren:resum}).

\begin{figure}[t]
\includegraphics[width=12cm]{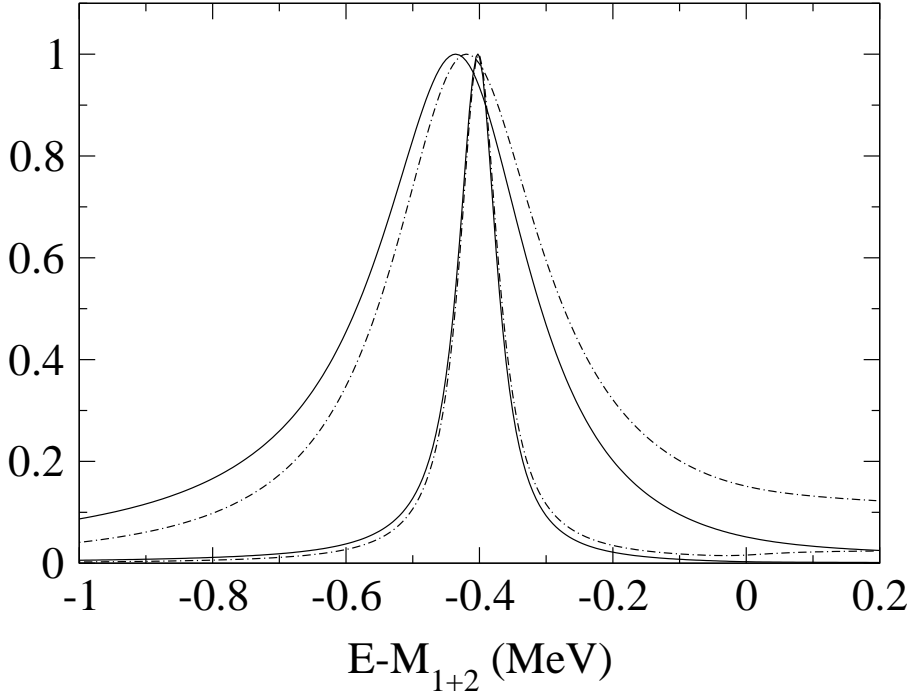}
\caption{
Line shapes of the $X$ resonance at NLO in the complex mass scheme.
The solid lines are for a short-distance decay channel
($|\mathcal{A}_2(E)|^2$ using $\mathcal{A}_2(E)$ in
Eq.~(\ref{A2-ren:resum})).
The dash-dotted lines are for the $D_1 D_1 \phi$ channel
(${\rm Im} \, \mathcal{A}_2(E)$ using $\mathcal{A}_2(E)$ in
Eq.~(\ref{A2-ren:resum})).
The line shapes are shown for $\kappa_X = 27.8$ MeV and
for both $\Gamma_2 = 0.07$ MeV (narrower resonances)
and $\Gamma_2 = 0.28$ MeV (wider resonances).
The line shapes are normalized so their maximum values are 1.
\label{fig:lineshapeAW}}
\end{figure}

In Fig.~\ref{fig:lineshapeAW}, we compare the line shape
of $X$ in the $D_1 D_1 \phi$ channel, which is proportional to
${\rm Im} \, \mathcal{A}_2(E)$,
with the line shape of $X$ in a short-distance decay channel,
which is proportional to $|\mathcal{A}_2(E)|^2$.
They are shown for two values of the
coupling constant $g$ that correspond to the width of $D_2$ being
$\Gamma_2 = 0.07$ MeV and $0.28$ MeV.
The curves are normalized so their maximum values are 1.
The line shapes have the same qualitative behavior.
The line shape for $D_1 D_1 \phi$ is shifted towards
higher energy by an amount that decreases as $g$ decreases.

The Belle Collaboration has observed an enhancement in the
production of $D^0 \bar D^0 \pi^0$ near threshold in the
decay $B \to K + D^0 \bar D^0 \pi^0$ \cite{Gokhroo:2006bt}.
The enhancement peaks at the invariant mass
$3875.2^{+1.1}_{-1.9}$ MeV, where we have combined the
errors in quadrature.  If we use the recent precision measurement
of the $D^0$ mass by the CLEO Collaboration \cite{Cawlfield:2007dw},
the peak is above the $D^{*0} \bar D^0$ threshold
by $3.4^{+1.2}_{-1.9}$ MeV.  In contrast, the peak observed
in the short-distance decay channel $J/\psi \, \pi^+ \pi^-$
is below the $D^{*0} \bar D^0$
threshold by $0.6 \pm 0.6$ MeV.  The difference is
$4.0^{+1.3}_{-2.0}$ MeV, which differs from 0 by two standard deviations.
Part of the discrepancy may be due to a shift in the
invariant mass distribution analogous to the one illustrated in
Fig.~\ref{fig:lineshapeAW}.

\subsection{Decay of $\bm{X}$ into $\bm{D_1D_1\phi}$}
\label{sec:Xdecay}

The width $\Gamma_X$ of the $X$ resonance was defined in
Eq.~(\ref{Epole}) in terms of the energy $E_{\rm pole}$
of the pole in the $X$ propagator:
$\Gamma_X = -2 \, {\rm Im} \, E_{\rm pole}$.
At LO in the complex mass scheme, $\Gamma_X$ is simply $\Gamma_2$.
At NLO in the complex mass scheme, $\Gamma_X$ is obtained by solving
Eq.~(\ref{gamma-pole:NLO}) for $E_{\rm pole}$.
If the width of the $X$ resonance is sufficiently small,
its width can also be interpreted as its decay rate.
A direct calculation of that decay rate will give a result
that is closely related to but not identical to $\Gamma_X$.
We will denote the result of the direct calculation of the
inclusive decay rate by $\Gamma[X \to D_1D_1\phi]$.

We can use the Optical Theorem to calculate the inclusive
decay rate for $X \to D_1D_1\phi$:
\begin{eqnarray}
\Gamma[X \to D_1 D_1 \phi] =
\frac{1}{M_X} \, {\rm Im} \, {\cal T}[X \to X].
\label{OpticalThm}
\end{eqnarray}
The forward T-matrix element for $X \to X$ can be defined using the LSZ
formalism and it is given in Eq.~(\ref{TXX}).  At NLO
in the complex mass scheme, the normalization
constant $Z_X$ and the propagator $\Delta_X(E,0)$
are given in Eqs.~(\ref{ZX-NLO}) and (\ref{DeltaX-2}).
The resulting expression for the decay rate of $X$
in Eq.~(\ref{OpticalThm}) is
\begin{eqnarray}
\Gamma[X \to D_1 D_1 \phi]
= -2 \, {\rm Im}
\frac{Z_X{\cal A}_2^{-1}(M_X)}{\lambda_0 [ L_0(M_X) + L_2(M_X)]}.
\label{ImTXX}
\end{eqnarray}
We have used Eq.~(\ref{A2-bare}) to express this in a form with a factor of
${\cal A}_2^{-1}(M_X)$, because ${\cal A}_2(E)$ does not depend
on the ultraviolet cutoff $\Lambda$.
The decay rate in Eq.~(\ref{ImTXX})
depends on $\Lambda$ through the
bare coupling constant $\lambda_0$ and through the ultraviolet-divergent
additive constants in $L_0(M_X)$ and $L_2(M_X)$.
From the renormalized expression for
${\cal A}_2(E)$ in Eq.~(\ref{A2-bare}), we can see that
the divergent terms in $L_0(E) + L_2(E)$ must be cancelled by $1/\lambda_0$
as $\Lambda \to \infty$.  The leading divergence is a linear divergence
in $L_0(M_{1+2})$ that is linear in $\Lambda$. Thus $1/\lambda_0$
must include a cancelling linear divergence. This implies that
$\lambda_0 [L_0(E) + L_2(E)]$ must approach 1 as $\Lambda \to \infty$.
Upon taking the limit $\Lambda \to \infty$, Eq.~(\ref{ImTXX}) reduces to
\begin{eqnarray}
\Gamma[X \to D_1 D_1 \phi] =
-2 \, {\rm Im} \left( Z_X \mathcal{A}_2^{-1}(M_X) \right).
\label{ImTXX-finite}
\end{eqnarray}

An explicit expression for the inclusive decay rate of $X$
can be obtained from Eq.~(\ref{ImTXX-finite}) by
inserting the renormalized expression for ${\cal A}_2(E)$
in Eq.~(\ref{A2-ren:resum}):
\begin{eqnarray}
\Gamma[X \to D_1 D_1 \phi] =
\frac{M_{12}}{2 \pi}  \, (-2) \,
{\rm Im} \left[ Z_X \big( \gamma(M_X) + F_{\rm cms}(\gamma(M_X))
    - {\rm Re} [\gamma_X + F_{\rm cms}(\gamma_X)] \big) \right],
\nonumber
\\
\label{GamDDphi}
\end{eqnarray}
where $Z_X$, $\gamma(E)$, $\gamma_X$, and $F_{\rm cms}(\gamma)$
are given in Eqs.~(\ref{ZX-NLO}), (\ref{gamma-E}), (\ref{gammaX}),
and (\ref{Fcms}). If we treat $\Gamma_2$ as order $g^2$,
the decay rate in Eq.~(\ref{GamDDphi}) reduces at order $g^2$
to the corresponding result for the width $\Gamma_X$ in
Eq.~(\ref{GamX:NLO}).

\begin{figure}[t]
\includegraphics[width=12cm]{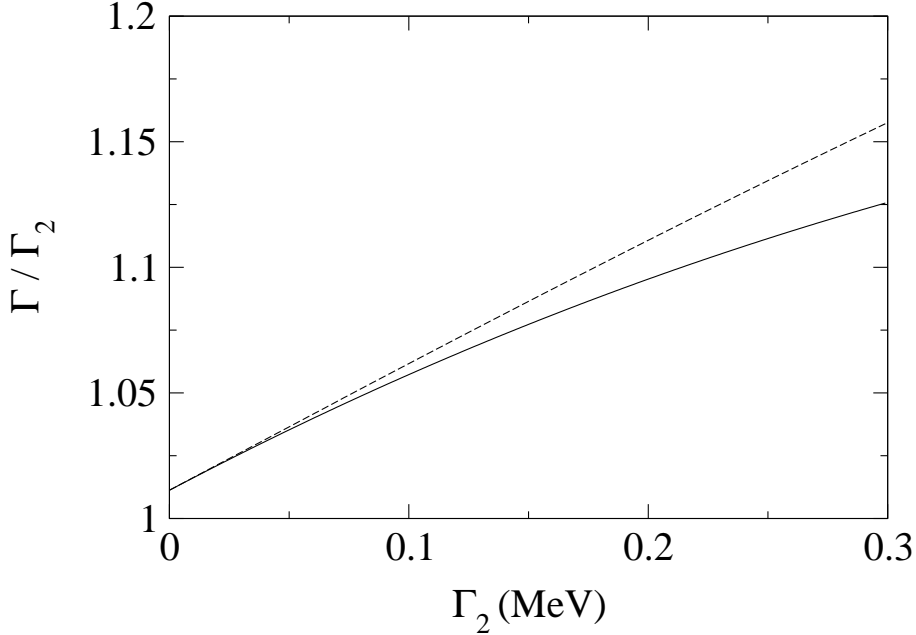}
\caption{ The width $\Gamma_X$ of the $X$ resonance (solid line) and
the decay rate $\Gamma[X \to D_1 D_1 \phi]$ (dashed line) at NLO in
the complex mass scheme as functions of $\Gamma_2$. The vertical
axis is in units of $\Gamma_2$.  The intercept on the vertical axis 
is approximately $1.011$.
\label{fig:widths}}
\end{figure}

In Fig.~\ref{fig:widths}, we compare the width $\Gamma_X$ obtained
by solving Eq.~(\ref{gamma-pole:NLO}) and the decay rate
$\Gamma[X \to D_1 D_1 \phi]$ in Eq.~(\ref{GamDDphi}) as functions of
$\Gamma_2$.  They agree for small $\Gamma_2$, because
the difference between $\Gamma_X$
and $\Gamma[X \to D_1 D_1 \phi]$ is of order $g^4$.
They differ by less than 1\% if $\Gamma_2 < 0.16$ MeV.

\section{Summary}
\label{sec:summary}

We have examined the effects of a weakly-bound hadronic molecule
on a nearby 3-body threshold, where the three bodies consist of a
primary constituent of the molecule and the decay products
of the other constituent.
We studied these effects in the simplest possible model:
the scalar meson model, with spin-0 constituents $D_1$, $D_2$,
and $\phi$ and momentum-independent interactions.
The primary constitutents of the molecule $X$ are $D_1 D_2$
and the 3-body threshold is that for $D_1 D_1 \phi$.
The $D_1 D_2$ contact interaction with coupling constant
$\lambda_0$ was treated nonperturbatively.
The $D_2-D_1 \phi$ interaction with coupling constant $g$ was
treated perturbatively.  Several observables were calculated to
all orders in $\lambda_0$ and to
next-to-leading order in $g$.  Both ultraviolet problems
and infrared problems were encountered in these calculations.

We found that at next-to-leading order in $g$, the ultraviolet
divergences can be removed by a perturbative renormalization
of the $D_2$ mass, a nonperturbative renormalization of $\lambda_0$,
and a nonperturbative renormalization of short-distance coefficients
in the operator product expansion.  The nonperturbative
renormalizations required the summing of a geometric series of
order-$g^2$ subdiagrams to all orders.  The subdiagrams
represent order-$g^2$ contributions to the amplitude for the
propagation of $D_1 D_2$ between contact interactions.
Thus the renormalized amplitudes at next-to-leading order
include all terms of order $g^2$ together with subsets of terms
from all higher orders in $g$.

We found that the next-to-leading order amplitudes suffer from
an infrared problem at the $D_1 D_2$ threshold.  The problem is
illustrated dramatically in Fig.~\ref{fig:lineshapeNLOcmsLO},
which shows the line shape $|\mathcal{A}(E)|^2$ in a
short-distance decay channel as a function of the energy $E$.
The pathological behavior at $E= M_{1+2}$ arises because the
$D_2-D_1 \phi$ interaction shifts the $D_1 D_2$
threshold from the real value $M_{1+2}$ to the complex value
$M_{1+2} - i \Gamma_2/2$.  Since $\Gamma_2$ is of order $g^2$,
a strict expansion in powers of $g$ leads to singularities at
$E= M_{1+2}$.  One possible solution to this problem is to sum
the geometric series of self-energy corrections
to the $D_2$ propagator to all orders,
but this leads to very complicated loop integrals.  A simpler
solution is to use the complex mass scheme, in which the width
$\Gamma_2$ is included in the Feynman rule for the $D_2$ propagator
and its effects are systematically compensated at higher orders
in $g$ through counterterms.  This partial resummation of the $D_2$
propagator corrections leads to smooth behavior of the line shape
$|\mathcal{A}(E)|^2$ at the $D_1 D_2$ threshold
as illustrated in Fig.~\ref{fig:lineshapeNLOcmsNLO}.
Having solved the infrared problems, we calculated several observables
to NLO in the complex mass scheme.  They include the line shapes
of the $X$ resonance in both a short-distance decay channel
and in the $D_1 D_1 \phi$ channel.

The $X(3872)$ is a weakly-bound hadronic molecule whose primary
constitutents are a superposition of charm mesons:
$D^{*0} \bar D^0 + D^0 \bar D^{*0}$.
Its mass is near the 3-body thresholds for $D^0 \bar D^0 \pi^0$,
$D^0 D^- \pi^+$, and $D^+ \bar D^0 \pi^-$.
The methods we applied to the scalar meson model can be extended
straightforwardly to this multichannel problem.
A conventional perturbation expansion in the $D^*-D\pi$ coupling constant
will have infrared problems at the $D^{*0} \bar D^0$ threshold
which are associated with the decay $D^{*0} \to  D^0 \pi^0$.
These infrared problems can be avoided by using the complex mass scheme.
The ultraviolet problems can be expected to be more severe in this
system, because the $D^*-D \pi$ interaction is proportional
to the 3-momentum of the pion.  Thus renormalization may not be
as simple as in the scalar meson model.
If the renormalization problem can be solved, it will be
straightforward to calculate observables to NLO in the complex mass
scheme.  One of the most interesting applications will be to predict the
difference between the line shapes of the $X(3872)$ resonance in
a short-distance decay channel, such as $J/\psi \, \pi^+ \pi^-$,
and in the $D^0 \bar D^0 \pi^0$ channel.

\begin{acknowledgments}
EB thanks S.~Fleming, M.~Kusunoki, and T.~Mehen for valuable discussions.
JL thanks the Physics Department at the Ohio State University
for its hospitality while some of this work was carried out.
This research was supported in part by the Department of Energy
under grant DE-FG02-91-ER4069 and by the Korea Research Foundation
under MOEHRD Basic Research Promotion grant KRF-2006-311-C00020.
\end{acknowledgments}


\appendix

\section{Loop Integrals}
\label{sec:loop}

In this appendix, we calculate the loop integrals that appear
at order $g^0$ and at order $g^2$ in the scalar meson model.

\subsection{Dimensional regularization}

Many of the loop integrals we need to evaluate are ultraviolet divergent.
They can be regularized by using dimensional regularization
of the integrals over the 3-momenta.  We analytically continue
the 3-dimensional integrals to $3-2\epsilon$ dimensions
using the prescription
\begin{equation}
\int_{\vec k} \equiv
\lim_{\epsilon \to 0} ~ (4 \pi)^{- \epsilon}
\left( \mbox{$\frac 3 2$} \right)_{-\epsilon}
\mu^{2 \epsilon}
\int \! \frac{d^{3-2 \epsilon} k}{(2 \pi)^{3-2 \epsilon}} ,
\label{dimreg}
\end{equation}
where $\mu$ is the renormalization scale
and $(z)_a$  is the Pochhammer symbol:
\begin{equation}
\left( z \right)_a = \frac{\Gamma(z+a)}{\Gamma(z)} .
\label{Pochhammer}
\end{equation}
The function of $\epsilon$ in the prefactor is designed to
simplify the analytic expressions for dimensionally regularized
integrals by cancelling the effects of the analytic continuation of
angular integrals. Thus the integral of a scalar function of $\vec
k$ is
\begin{equation}
\int_{\vec k} f(|\vec k \,|) \equiv
\lim_{\epsilon \to 0} ~
\frac{1}{2\pi^2} \mu^{2 \epsilon}
\int_0^\infty\!dk~ k^{2-2\epsilon} f(k) .
\label{dimreg-scalar}
\end{equation}

Loop integrals are evaluated by first inserting the appropriate
nonrelativistic propagators:
\begin{eqnarray}
\Delta_i(p_0,p) &=&
\frac{i}{p_0 - M_i - p^2/(2M_i) + i \varepsilon} ,
\label{Delta-NR}
\end{eqnarray}
where $i=1,2,\phi$ and $M_\phi = m$.
Integrals over the loop energy are then evaluated using
the residue theorem.  After using the Feynman parameter trick
to combine denominators, the dimensionally regularized integrals
over the 3-momenta can be evaluated.
The final steps are the evaluation of the
Feynman parameter integrals and the analytic continuation
to $\epsilon = 0$.
It will be convenient to express the results in terms of the
variables
\begin{subequations}
\begin{eqnarray}
\kappa &=& \left[ -2 M_{12} (E - M_{1+2}) - i \varepsilon \right]^{1/2},
\label{kappa-def}
\\
\kappa_1 &=& \left[ 2 M_{12} (M_{2-1} - m) \right]^{1/2}.
\end{eqnarray}
\label{kappakappa1}
\end{subequations}

\subsection{Self-energy of $\bm{D_2}$}
\label{sec:Self-energy}

The one-loop diagram for the self-energy $\Sigma(p_0,p)$
of $D_2$ is shown in Fig.~\ref{fig:self}(a).
The expression for the diagram is
\begin{equation}
\Sigma_{2a}(p_0,p) =
-i g^2 \int \! \! \frac{d^4q}{(2 \pi)^4} \,
\Delta_1(q_0,q) \, \Delta_\phi(p_0-q_0,|\vec p - \vec q \,|) .
\label{Sigma2-1loop}
\end{equation}
The residue theorem can be used
to evaluate the energy integral:
\begin{equation}
\Sigma_{2a}(p_0,p) =
g^2\int_{\vec q}
\frac{1}{p_0 - M_1 - m - q^2/(2M_1)
    - (\vec p - \vec q \,)^2/(2m) + i \varepsilon}  .
\label{Sigma2a}
\end{equation}
If we implement dimensional
regularization using the prescription in Eq.~(\ref{dimreg}), the
analytic result is
\begin{equation}
\Sigma_{2a}(p_0,p) =
\frac{g^2 m_1 \mu^{2 \epsilon}}{2\pi \cos (\pi \epsilon)}
\left( - 2 m_1 \bigg[ p_0 - M_1 - m - \frac{p^2}{2(M_1+m)}\bigg]
        - i \varepsilon  \right)^{\frac{1}{2} - \epsilon} .
\label{Sigma2-dimreg}
\end{equation}
The self-energy in dimensional regularization is given by
the analytic continuation of
Eq.~(\ref{Sigma2-dimreg}) to $\epsilon = 0$:
\begin{equation}
\Sigma_{2a}(p_0,p) =
\frac{g^2 m_1}{2 \pi}
\left( - 2 m_1 \bigg[ p_0 - M_1 - m - \frac{p^2}{2(M_1+m)}\bigg]
        - i \varepsilon  \right)^{\frac{1}{2}} .
\label{Sigma2-dimreg0}
\end{equation}

The momentum integral in Eq.~(\ref{Sigma2a}) has a linear ultraviolet
divergence that is set to zero by dimensional regularization.
The self-energy in a general regularization scheme is given in
Eq.~(\ref{D2self}).  The linear ultraviolet divergence is contained
in the extra term $\Sigma_{2a}(M_1+m,0)$, which is real valued.

\subsection{Propagation between contact interactions at leading order}
\label{sec:Propcon0}

The amplitude for the propagation of $D_1 D_2$
between contact interactions is given at order $g^0$
by the one-loop diagram in Fig.~\ref{fig:L-LO}.
In the center-of-momentum frame
where the 4-momentum is $(E,0)$, this amplitude is
\begin{equation}
i L_0(E) = \int \frac{d^4q}{(2 \pi)^4} \,
\Delta_2(q_0,q) \, \Delta_1(E-q_0,q) .
\end{equation}
The residue theorem can be used to evaluate the energy integral:
\begin{equation}
L_0(E) =
\int_{\vec q} \frac{1}{(E - M_{1+2}) - q^2/2M_{12}+ i \varepsilon} .
\label{L0-E}
\end{equation}
If we implement dimensional
regularization using the prescription in Eq.~(\ref{dimreg}), the
analytic result is
\begin{equation}
L_0(E) = \frac{M_{12}\mu^{2 \epsilon}}
    {2 \pi \cos (\epsilon \pi)}
\kappa^{1 - 2\epsilon} ,
\end{equation}
where $\kappa$ is defined in Eq.~(\ref{kappa-def}).
The amplitude in dimensional regularization is given by
analytic continuation to $\epsilon = 0$:
\begin{equation}
L_0(E) = \frac{M_{12}}{2 \pi } \kappa .
\label{L0-3d}
\end{equation}

The momentum integral in Eq.~(\ref{L0-E}) has a linear ultraviolet
divergence that is set to zero by dimensional regularization.
The amplitude in a general regularization scheme is given in
Eq.~(\ref{L0-sub}).  The linear ultraviolet divergence is contained
in the extra term $L_0(M_{1+2})$, which is real valued.

\subsection{Propagation between contact interactions at next-to-leading order}
\label{sec:Propcon2}

The amplitude for the propagation of $D_1 D_2$
between contact interactions has contribution of order $g^2$
from the two-loop diagrams in Fig.~\ref{fig:L-NLO}.
In the center-of-momentum frame where the 4-momentum is $(E,\vec 0)$,
the order-$g^2$ amplitude $i L_2(E)$ is a function of $E$ only.
We proceed to calculate the contributions from each of the
three diagrams.

\subsubsection{Correction from $\bm{\phi}$ exchange}

The contribution to the
amplitude $i L_2(E)$ from the two-loop diagram in Fig.~\ref{fig:L-NLO}(a)
in which $\phi$ is exchanged between the $D_1$ and $D_2$ is
\begin{eqnarray}
i L_{2a}(E) &=& - g^2 \int \frac{d^4k}{(2
\pi)^4} \int \frac{d^4q}{(2 \pi)^4}\, \Delta_2(k_0,k) \,
\Delta_1(E-k_0,k)
\nonumber
\\
&&\hspace{1cm} \times \Delta_2(q_0,q) \, \Delta_1(E-q_0,q) \,
\Delta_\phi(k_0 + q_0 - E,|\vec k + \vec q \,|) .
\end{eqnarray}
The integrals over the loop energies $k_0$ and $q_0$ can be
evaluated by closing the contours in the upper half-planes.
After introducing Feynman parameters and evaluating the
integrals over the loop momenta, the amplitude reduces to
\begin{eqnarray}
L_{2a}(E) &=&
- \frac{g^2 M_{12}^2 m_1}{16 \pi^3} \,
\frac{(1-2 \epsilon)^2 \left( \frac 1 2 \right)^2_{-\epsilon}
    \left( 1 \right)_{\epsilon}}{\epsilon}
\mu^{4 \epsilon}
\nonumber
\\
&& \times
\int_0^1 \! dx \int_0^{1-x} \! dy \,
\left[ (1-x)(1-y) - (m_1/m)^2 (1 - x - y)^2 \right]^{-3/2 + \epsilon}
\nonumber
\\
&& \hspace{2cm}
\times
\left[ (x+y) \kappa^2 + (1-x-y)(m_1/M_{12})(\kappa^2-\kappa_1^2)\right]^{-2 \epsilon} ,
\end{eqnarray}
where $\kappa$ and $\kappa_1$ are defined in Eqs.~(\ref{kappakappa1}).
One of the Feynman parameter integrals can be evaluated analytically
by changing variables to $z = x+y$ and $w= (x-y)/(x+y)$.
The integral over $w$ gives a hypergeometric function:
\begin{eqnarray}
L_{2a}(E) &=&
- \frac{g^2 M_{12}^2m_1}{2 \pi^3} \,
\frac{(1-2 \epsilon)^2 \left( \frac 1 2 \right)^2_{-\epsilon}
    \left( 1 \right)_{\epsilon}2^{-2 \epsilon}}{\epsilon}
\mu^{4 \epsilon}
\int_0^1 \! dz \, z^{-2 + 2 \epsilon} \, t(z)^{3/2-\epsilon} \,
\nonumber
\\
&& \hspace{1cm} \times
{}_2F_1(\mbox{$\frac{3}{2} - \epsilon$},\mbox{$\frac{1}{2}$},
    \mbox{$\frac{3}{2}$};t(z))
\left[ z \kappa^2 + (1-z)(m_1/M_{12})(\kappa^2 - \kappa_1^2)\right]^{-2 \epsilon} ,
\label{L3-eps}
\end{eqnarray}
where $t(z)$ is the rational function of $z$ given in Eq.~(\ref{t-z}).
The pole at $\epsilon = 0$ in Eq.~(\ref{L3-eps})
is a logarithmic ultraviolet divergence.

The amplitude $L_{2a}(E)$ in dimensional regularization is
defined by the Laurent expansion of the expression in
Eq.~(\ref{L3-eps}) to order $\epsilon^0$.
The expansion in powers of $\epsilon$ is facilitated by
using a transformation formula to replace the
hypergeometric function in Eq.~(\ref{L3-eps}) by one of the form
$_2F_1(1,\epsilon,\frac{3}{2};t)$:
\begin{equation}
{}_2F_1(\mbox{$\frac{3}{2} - \epsilon$},\mbox{$\frac{1}{2}$},
    \mbox{$\frac{3}{2}$};t) =
{}_2F_1(1,\epsilon,\mbox{$\frac{3}{2}$};t)
(1 - t)^{-1/2 + \epsilon} .
\end{equation}
The expansion of $_2F_1(1,\epsilon,\frac{3}{2};t)$
to order $\epsilon$ is simple:
\begin{equation}
{}_2F_1(1,\epsilon,\mbox{$\frac{3}{2}$};t) = 1 +
\mbox{$\frac{2}{3}$} \epsilon \, t \; {}_2F_1(1,1,\mbox{$\frac{5}{2}$};t) + \ldots .
\end{equation}
Expanding the amplitude $L_{2a}(E)$ in Eq.~(\ref{L3-eps})
to 0$^{\rm th}$ order in $\epsilon$, it reduces to
\begin{eqnarray}
L_{2a}(E) &=&
- \frac{g^2 M_{12}^2 m}{2 \pi^3} \,
\Bigg( \frac{1}{2} \arcsin\frac{m_1}{m}
\left( \frac{1}{\epsilon} - 4 + \gamma + 2 \ln 2 \right)
\nonumber
\\
&& \hspace{1cm}
+ \frac{m_1}{m} \int_0^1 \frac{dz}{z^2} \,
\left( \frac{t^3(z)}{1-t(z)} \right)^{1/2}
\bigg[ \ln \frac{z^2 [1-t(z)]}{t(z)}
+ \frac{2t(z)}{3} \,
{}_2F_1 \left( 1,1,\mbox{$\frac{5}{2}$};t(z) \right)
\nonumber
\\
&& \hspace{5cm}
-2 \ln \frac{z\kappa^2+(1-z)(m_1/M_{12})(\kappa^2-\kappa_1^2)}{\mu^2}
\bigg] \Bigg).
\label{L3-dimreg}
\end{eqnarray}
We have used the integral
\begin{equation}
\int_0^1 \frac{dz}{z^2} \,
\left( \frac{t^3(z)}{1-t(z)} \right)^{1/2} =
\frac{m}{2 m_1} \arcsin \frac{m_1}{m}.
\end{equation}
This can be evaluated by changing variables
to $w = z/[z + 2r(1-z)]$.
Most of the terms in Eq.~(\ref{L3-dimreg}) are independent of $E$.
The expression can be simplified by expressing it in the form
\begin{eqnarray}
L_{2a}(E) = L_{2a}(2M_1 + m) &+&
\frac{g^2 M_{12}^2 m_1}{\pi^3}
\int_0^1 \frac{dz}{z^2}
\left( \frac{t^3(z)}{1-t(z)} \right)^{1/2}
\nonumber
\\
&& \hspace{2cm}
\times
\ln \frac{z \kappa^2 + (1-z) (m_1/M_{12}) (\kappa^2 - \kappa_1^2)}
    {z \kappa_1^2} .
\label{L3-integral}
\end{eqnarray}
The amplitude $L_{2a}(2M_1 + m)$ at the $D_1 D_1 \phi$ threshold
is real-valued and includes the logarithmic divergence.
The amplitude $L_{2a}(E)$ has an imaginary part for $E > 2 M_1 + m$.
For energies in the range $2 M_1 + m < E < M_{1+2}$,
the imaginary part is
\begin{eqnarray}
{\rm Im} \, L_{2a}(E)  &=&
-\frac{g^2 M_{12}^2m_1}{\pi^2}
\int_0^{z_0} \frac{dz}{z^2}
\left( \frac{t^3(z)}{1-t(z)} \right)^{1/2} ,
\label{ImL2a}
\end{eqnarray}
where the upper limit $z_0$ of the integral is
\begin{equation}
z_0 =
\frac{(m_1/M_{12})(\kappa_1^2 - \kappa^2)}
    {(m_1/M_{12})(\kappa_1^2 - \kappa^2) + \kappa^2}.
\label{z0}
\end{equation}
The first few terms in the Laurent expansion of $L_{2a}(E)$
in powers of $\kappa$ are
\begin{eqnarray}
L_{2a}(E) \approx L_{2a}(2M_1 + m) &+&
\frac{g^2 M_{12}^2 m_1}{2 \pi^3}
\Bigg[ 2 \int_0^1 \frac{dz}{z^2}
\left( \frac{t^3(z)}{1-t(z)} \right)^{1/2}
    \ln \frac{(1-z)m_1}{zM_{12}}
\nonumber
\\
&& \hspace{1.5cm}
- i \pi \frac{m}{m_1} \arcsin \frac{m_1}{m}
+ 2 \pi i \left( \frac{M_{12}}{m_1} \right)^{1/2} \frac{\kappa}{\kappa_1} \Bigg] .
\label{L2a-Laurent}
\end{eqnarray}

\subsubsection{Correction from $\bm{D_2}$ self-energy}
\label{app:D2self}

The contribution to $i L_2(E)$
from the two-loop diagram in Fig.~\ref{fig:L-NLO}(b)
with a self-energy correction to the $D_2$ line is
\begin{eqnarray}
i L_{2b}(E) &=& -
i \int \frac{d^4k}{(2 \pi)^4} \,
\Delta_2(k_0,k)^2 \, \Delta_1(E-k_0,k) \, \Sigma_{2a}(k_0,k)  ,
\label{L2b}
\end{eqnarray}
where $\Sigma_{2a}(k_0,k)$ is the contribution to the $D_2$
self-energy from the one-loop diagram in Fig.~\ref{fig:self}(a).
The analytic result for $\Sigma_{2a}(k_0,k)$
in dimensional regularization is given in Eq.~(\ref{Sigma2-dimreg}).
Since $\Sigma_{2a}(k_0,k)$ has a branch cut for $k_0$ in the lower
half-plane, it is convenient to close the $k_0$ contour in the upper
half-plane. After introducing Feynman parameters, the integral over
$\vec k$ can be evaluated. If we implement dimensional regularization
using the prescription in Eq.~(\ref{dimreg}), the result is
\begin{eqnarray}
L_{2b}(E) &=&
- \frac{g^2 M_{12}^2m_1}{16 \pi^3} \,
\left( \frac{m_1}{M_{11}} \right)^{1/2-\epsilon} \,
\frac{(1-2 \epsilon)^2 (1)_{2\epsilon}}
    {( \frac 1 2 )_{\epsilon}^2 \cos^2(\pi \epsilon) \epsilon} \mu^{4 \epsilon}
\nonumber
\\
&&
\times
\int_0^1 \! dx \, x (1-x)^{-3/2+\epsilon}
\left[ x \kappa^2 + (1-x)(M_{11}/M_{12})(\kappa^2 - \kappa_1^2)
    \right]^{-2\epsilon},
\label{L2-eps}
\end{eqnarray}
where $\kappa$ and $\kappa_1$ are defined in Eqs.~(\ref{kappakappa1})
and $M_{11}$ is given by
\begin{eqnarray}
M_{11} &=& \frac{M_1(M_1+m)}{2M_1+m} .
\end{eqnarray}
The pole at $\epsilon = 0$ in Eq.~(\ref{L2-eps}) is a
logarithmic ultraviolet divergence.

The amplitude $L_{2b}(E)$ in dimensional regularization is
defined by the Laurent expansion of the expression in
Eq.~(\ref{L2-eps}) to order $\epsilon^0$.
Expanding to 0$^{\rm th}$ order in $\epsilon$,
the expression reduces to
\begin{eqnarray}
L_{2b}(E) &=&
\frac{g^2 M_{12}^2 m_1}{4 \pi^3}
\left( \frac{m_1}{M_{11}} \right)^{1/2} \,
\Bigg[ \frac{1}{\epsilon} - 4 + 4\ln2 - \ln \frac{m_1}{M_{11}}
- 2 \ln \frac{\kappa^2}{\mu^2}
\nonumber
\\
&& \hspace{0cm}
+ \mbox{$\frac{1}{2}$} \int_0^1 \! dx \, x (1-x)^{-3/2}
\ln \frac{x \kappa^2 + (1-x)(M_{11}/M_{12})(\kappa^2 - \kappa_1^2)}
        {\kappa^2} \Bigg] .
\label{L2-dimreg}
\end{eqnarray}
The integral over $x$ can be evaluated analytically.
We can simplify the expression by replacing $M_{11}/M_{12}$ by 1,
because the difference is suppressed by a
factor of $\kappa_1^2/M_2^2$.
Most of the terms in Eq.~(\ref{L2-dimreg}) are independent of $E$.
The expression can be simplified by expressing it in the form
\begin{eqnarray}
L_{2b}(E) = L_{2b}(2M_1 + m)
- \frac{g^2 M_{12}^2 m_1}{2 \pi^3}
\left( \frac{m_1}{M_{12}} \right)^{1/2} \,
\left( 2 \ln \frac{\kappa+\kappa_1}{2 \kappa_1}
+ \frac{(\kappa - \kappa_1)^2}{2 \kappa \kappa_1}
\ln \frac{\kappa + \kappa_1}{\kappa - \kappa_1} \right).
\nonumber
\\
&&
\label{L2-analytic}
\end{eqnarray}
The amplitude $L_{2b}(2M_1 + m)$ at the $D_1 D_1 \phi$ threshold
is real-valued and includes the logarithmic divergence.
The amplitude $L_{2b}(E)$ has an imaginary part for $E > 2 M_1 + m$.
For energies in the range $2 M_1 + m < E < M_{1+2}$,
the imaginary part of $L_{2b}(E)$ is
\begin{eqnarray}
{\rm Im} \, L_{2b}(E) &=&
- \frac{g^2 M_{12}^2 m_1}{4 \pi^2}
\left( \frac{m_1}{M_{12}} \right)^{1/2} \,
\frac{(\kappa_1 - \kappa)^2}{\kappa \kappa_1} .
\label{ImL2b}
\end{eqnarray}
Note that this diverges at the $D_1 D_2$ threshold $\kappa = 0$.
The first few terms in the Laurent expansion of $L_{2a}(E)$
in powers of $\kappa$ are
\begin{eqnarray}
L_{2b}(E) \approx L_{2b}(2M_1 + m)
- \frac{g^2 M_{12}^2 m_1}{2 \pi^3}
\left( \frac{m_1}{M_{12}} \right)^{1/2}
\left[ 1 - 2 \ln 2 + i \pi
\left( \frac{\kappa_1}{2 \kappa} - 1
    + \frac{\kappa}{2 \kappa_1} \right) \right] .
\nonumber
\\
\label{L2b-Laurent}
\end{eqnarray}

\subsubsection{Correction from counterterms}

In a general regularization scheme, the one-loop self-energy
subdiagram $\Sigma_{2a}(k_0,k)$ in Eq.~(\ref{L2b})
has a linear ultraviolet divergence, which is included in the term
$\Sigma_{2a}(M_1+m,0)$ in Eq.~(\ref{D2self}).
The resulting divergence in $L_{2b}(E)$ is cancelled by the
one-loop diagram with a $D_2$ mass
counterterm in Fig.~\ref{fig:L-NLO}(c).
The counterterm vertex is $- i \delta M$,
with $\delta M = \Sigma_{2a}(M_1+m,0)$.

In the complex mass scheme, the $D_2$ propagator counterterm
has the more complicated form given in Eq.~(\ref{newcounterterm}).
With this counterterm, the contribution to $L_2(E)$ of order $g^2$
from the one-loop diagram in Fig.~\ref{fig:L-NLO}(c) is
\begin{eqnarray}
i L_{2c}(E) &=&
-i \int \frac{d^4k}{(2\pi)^4}
\Delta_1(E-k_0,k) \Delta_2(k_0,k)^2
\Big( (\delta M + i \Gamma_2/2)
\nonumber
\\
&& \hspace{5cm}
    + (Z_2^{-1}-1) [ k_0 - M_2 -k^2/(2M_2) ] \Big).
\end{eqnarray}
In the term with the factor $Z_2^{-1}-1$, the integral
is proportional to $L_0(E)$. This amplitude in a general
regularization scheme is given in Eq.~(\ref{L0-sub}).
In the term with the factor $\delta M + i \Gamma_2/2$,
the integral is convergent. The complete result for the diagram is
\begin{eqnarray}
L_{2c}(E) =
(\delta M + i \Gamma_2/2) \frac{M_{12}^2}{2\pi \kappa}
+ (Z_2^{-1}-1) \left( L_0(M_{1+2}) + \frac{M_{12}}{2\pi}\kappa \right) .
\label{L2c}
\end{eqnarray}


\end{document}